\documentclass{revtex4}
\usepackage{graphicx}
\usepackage{fancyhdr}
\usepackage{amsmath}
\newcommand{\nn}{\nonumber}
\newcommand\sss{\scriptscriptstyle}

\usepackage{color}

\begin{document}

\title{Characterising a Higgs-like resonance at the LHC}

%

\author{Priscila de Aquino and Kentarou Mawatari}
\thanks{speaker}
\email{kentarou.mawatari@vub.ac.be}
\affiliation{Theoretische Natuurkunde and IIHE/ELEM, Vrije
  Universiteit Brussel\\ and International Solvay Institutes,
  Brussels, Belgium}

\begin{abstract}
 We present the implementation of an effective lagrangian via 
 {\sc FeynRules}, featuring bosons $X(J^P)$ with various assignments of
 spin/parity $J^P=0^+$, $0^-$, $1^+$, $1^-$, or $2^+$, that allows one
 to perform characterisation studies of the boson recently discovered at
 the LHC,
 for all the relevant channels and in a
  consistent, systematic and accurate way.
\end{abstract}

\maketitle

\thispagestyle{fancy}


\section{Introduction}

The recent observation of a new boson with a mass of about 125~GeV at the
LHC~\cite{Aad:2012tfa} brought us a lot of
excitement. While this discovery is a great triumph for theoretical and
experimental high energy physics, the detailed study of the new state
will require many years of work at the LHC as well as at the ILC. The study
includes the determination of its spin and parity quantum numbers and
the coupling strength for the interactions, which can tell us if the
observed resonance is indeed responsible for the Brout-Englert-Higgs
mechanism~\cite{Englert:1964et}, or involves physics beyond
the Standard Model (SM).  
 
We introduce a complete framework, based on an effective filed theory
approach, that allows one to perform characterisation studies of the
recently-discovered boson. 
Our assumptions are simply that the resonance structure
observed in data corresponds to one bosonic state ($X(J^P)$ with
$J=0^+$, $0^-$, $1^+$, $1^-$,
or $2^+$, and a mass of about $125$ GeV), and that no other new state below the
cutoff $\Lambda$ coupled to such a resonance exists. We also follow the
principle that any new physics is dominantly described by the lowest
dimensional operators.
This means, for example, that for the spin-0 $CP$-even
case (which corresponds to the SM scalar) we include all effects coming from
the complete set of dimension-six operators relevant to Higgs
observables. Given that our goal is that of providing a simulation framework
in terms of mass eigenstates, and consistently with the general guidelines
outlined above, we construct an effective lagrangian below the EWSB
scale, where $SU(2)_L \times U(1)_Y$ reduces to $U(1)_{EM}$; moreover, we do
not require $CP$ conservation, and we leave open the possibility that the new
boson might be a scalar with no definite $CP$ properties.

Technically, the implementation of the lagrangian is
performed in {\sc FeynRules}~\cite{Christensen:2008py} extending and
completing the earlier version used in ref.~\cite{Englert:2012xt}. The
particle content and the Feynman rules of the model can be exported to
any matrix element generator in the UFO
format~\cite{Degrande:2011ua}. We dub it 
\emph{Higgs Characterisation model}~\cite{Artoisenet:2013puc} 
and it can be found on the {\sc FeynRules} on-line 
database at {\tt http://feynrules.irmp.ucl.ac.be}.   

There are several advantages in having a first principle implementation
in terms of an effective lagrangian which can be automatically
interfaced to a matrix element generator (and then to an event
generator). First and most important, all relevant production and decay
modes can be studied within the same model, from gluon fusion to VBF as
well as $VH$ and $t\bar tH$ associated productions can be considered and
the corresponding processes automatically generated within
minutes. Second, it is straightforward to modify the model
implementation to extend it further in case of need, by adding further
interactions, for example of higher-dimensions. Finally, higher-order
QCD effects can be easily accounted for by multi-parton tree-level or full
NLO computations and their matching with parton showers in
automatic frameworks,
e.g. with {\sc MadGraph5}~\cite{Alwall:2011uj} or 
{\sc aMC@NLO}~\cite{Frederix:2009yq}.

In this report we first write down the effective lagrangian explicitly,
and then show mass and angular distributions in the $pp\to X(\to
ZZ^*/WW^*)\to 4\ell$ process, 
comparing with the results by the JHU program~\cite{Bolognesi:2012mm}
which takes the anomalous coupling framework and is
employed by both ATLAS and CMS collaborations. 
See ref.~\cite{Artoisenet:2013puc} for all the detailed demonstration and analyses. 

\section{The effective lagrangian}

\subsection{Spin 0}

The construction of the effective lagrangian for the spin-0 state is obtained
by requiring that the parametrisation: i) allows one to recover the SM case
easily; ii) includes all possible interactions that are generated by
gauge-invariant dimension-six operators above the EW scale; iii) includes
$0^-$ state couplings typical of SUSY or of generic two-Higgs-doublet models
(2HDM); and iv) allows $CP$-mixing between $0^+$ and $0^-$ states (which we
parametrise in terms of an angle $\alpha$).

Let us start with the interaction lagrangian relevant to fermions which,
while being extremely simple, illustrates our philosophy well. 
Such a lagrangian is:
\begin{align}
 {\cal L}_0^f 
   = -\sum_{f=t,b,\tau}\bar\psi_f\big( c_{\alpha}\kappa_{\sss Hff}g_{\sss Hff}\,
        +i s_{\alpha}\kappa_{\sss Aff}g_{\sss Aff}\, \gamma_5 \big)
\psi_f X_0 \,,
\label{eq:0ff}
\end{align}
where we use the notation $c_\alpha\equiv \cos \alpha$ and
$s_\alpha\equiv \sin \alpha$, and 
denote by $g_{Hff}=m_f/v$ $(g_{Aff}=m_f/v)$ the strength of the scalar 
(pseudoscalar) coupling in the SM (in a 2HDM with $\tan\beta=1$). We point 
out that the constants $\kappa_i$ can be taken real without any loss of
generality. For simplicity, we have assumed that only the third-generation
of fermions couple to the scalar state; extensions to the other families and
flavour-changing structures are trivial to implement, which can be directly 
done by users of {\sc FeynRules}.  
The interaction of
eq.~(\ref{eq:0ff}) can also parametrise the effects of a ${\cal L}^{\rm
  dim=6}_Y = (\phi^\dagger \phi) Q_L \tilde \phi t_R$ operator, which
  modifies the value of the Yukawa coupling, but not the interaction structure.
Note also that all requirements listed above are satisfied at the
price of a small redundancy in the number of parameters. The SM is obtained
when $c_{\alpha}=1$ and $\kappa_{Hff}=1$. The pseudoscalar state of a type-II
$CP$-conserving 2HDM or SUSY is obtained by setting $s_{\alpha}=1$ and
$\kappa_{Aff}=\cot \beta$ or $\kappa_{Aff}=\tan \beta$ for up or down
components of the $SU(2)$ fermion doublet, respectively. The parametrisation
of $CP$ mixing is entirely realised in terms of the angle $\alpha$, i.e.
independently of the parameters $\kappa_i$, so that many interesting cases,
such as again $CP$-violation in generic 2HDM, can be covered.

The effective lagrangian for the interaction of scalar and pseudoscalar states
with vector bosons can be written as follows:
\begin{align}
 {\cal L}_0^V =\bigg\{&
  c_{\alpha}\kappa_{\rm SM}\big[\frac{1}{2}g_{\sss HZZ}\, Z_\mu Z^\mu 
                                +g_{\sss HWW}\, W^+_\mu W^{-\mu}\big] \nn\\
  &-\frac{1}{4}\big[c_{\alpha}\kappa_{\sss H\gamma\gamma}
 g_{\sss H\gamma\gamma} \, A_{\mu\nu}A^{\mu\nu}
        +s_{\alpha}\kappa_{\sss A\gamma\gamma}g_{ \sss A\gamma\gamma}\,
 A_{\mu\nu}\widetilde A^{\mu\nu}
 \big] 
  -\frac{1}{2}\big[c_{\alpha}\kappa_{\sss HZ\gamma}g_{\sss HZ\gamma} \, 
 Z_{\mu\nu}A^{\mu\nu}
        +s_{\alpha}\kappa_{\sss AZ\gamma}g_{\sss AZ\gamma}\,Z_{\mu\nu}\widetilde A^{\mu\nu} \big] \nn\\
  &-\frac{1}{4}\big[c_{\alpha}\kappa_{\sss Hgg}g_{\sss Hgg} \, G_{\mu\nu}^aG^{a,\mu\nu}
        +s_{\alpha}\kappa_{\sss Agg}g_{\sss Agg}\,G_{\mu\nu}^a\widetilde G^{a,\mu\nu} \big] \nn\\
  &-\frac{1}{4}\frac{1}{\Lambda}\big[c_{\alpha}\kappa_{\sss HZZ} \, Z_{\mu\nu}Z^{\mu\nu}
        +s_{\alpha}\kappa_{\sss AZZ}\,Z_{\mu\nu}\widetilde Z^{\mu\nu} \big] 
  -\frac{1}{2}\frac{1}{\Lambda}\big[c_{\alpha}\kappa_{\sss HWW} \, W^+_{\mu\nu}W^{-\mu\nu}
        +s_{\alpha}\kappa_{\sss AWW}\,W^+_{\mu\nu}\widetilde W^{-\mu\nu}\big] \nn\\ 
  &-\frac{1}{\Lambda}c_{\alpha} 
    \big[ \kappa_{\sss H\partial\gamma} \, Z_{\nu}\partial_{\mu}A^{\mu\nu}
         +\kappa_{\sss H\partial Z} \, Z_{\nu}\partial_{\mu}Z^{\mu\nu}
         +\kappa_{\sss H\partial W} \, \big(W_{\nu}^+\partial_{\mu}W^{-\mu\nu}+h.c.\big)
 \big]
 \bigg\} X_0\,,
 \label{eq:0vv}
\end{align}
where the (reduced) field strength tensors are defined as follows:
\begin{align}
 V_{\mu\nu} =\partial_{\mu}V_{\nu}-\partial_{\nu}V_{\mu}\quad (V=A,Z,W^{\pm})\,,\quad 
 G_{\mu\nu}^a =\partial_{\mu}^{}G_{\nu}^a-\partial_{\nu}^{}G_{\mu}^a
  +g_sf^{abc}G_{\mu}^bG_{\nu}^c\,,
\end{align}
and the dual tensor is $\widetilde V_{\mu\nu} =\frac{1}{2}\epsilon_{\mu\nu\rho\sigma}V^{\rho\sigma}$.
The parametrisation of the couplings to vectors follows the same principles as
that of the couplings to fermions. In particular, the mixing angle $\alpha$
allows for a completely general description of $CP$-mixed states. We stress
here that while in general in a given model $CP$ violation depends on the
whole set of possible interactions among the physical states and cannot be
established by looking only at a sub sector~\cite{CPV}, in our parametrisation
$\alpha\neq 0$ or $\alpha\neq\pi/2$ (and non-vanishing $\kappa_{Hff},
\kappa_{Aff},\kappa_{HVV},\kappa_{AVV}$) implies $CP$ violation.  This can be
easily understood by first noting that in eq.~(\ref{eq:0ff}) $\alpha \neq 0$
or $\alpha\neq\pi/2$ always leads to $CP$ violation and that the corresponding 
terms in eq.~(\ref{eq:0vv}) are generated via a fermion loop by the 
$X_0 ff$ interaction. The $CP$-odd analogues of the operators in the last 
line of eq.~(\ref{eq:0vv}) do vanish. 

In our implementation, the parameters listed in table~\ref{tab:param} can be
directly set by the user.  
The dimensionful couplings $g_{Xyy'}$ are
set so as to reproduce a SM Higgs and a pseudoscalar one in a 2HDM with
$\tan\beta=1$, e.g. $g_{HZZ}=2m_Z^2/v$ as well as
$g_{Hgg}=-\alpha_s/3\pi v$ and $g_{Agg}=-\alpha_s/2\pi v$ in the heavy top loop limit.

\begin{table}
\center
\begin{small}
\caption{Model parameters.}
\begin{tabular}{lll}
\hline
 parameter\hspace*{5mm} & reference value\hspace*{5mm} & description \\
\hline
 $\Lambda$ [GeV] & $10^3$ & cutoff scale \\
 $c_{\alpha}(\equiv \cos\alpha$) & 1 & mixing between $0^+$ and
         $0^-$ \\
 $\kappa_i$ & 0 , 1 & dimensionless coupling parameter \\
\hline
\end{tabular}
\label{tab:param}
\end{small}
\end{table}

\subsection{Spin 1}

The interaction Lagrangian for the spin-1 boson with fermions is written as
\begin{align}
 {\cal L}_1^f = \sum_{f=q,\ell} 
      \bar\psi_f \gamma_{\mu}(\kappa_{f_a}a_f - \kappa_{f_b}
           b_f\gamma_5)\psi_f X_{1}^{\mu}\,.
\end{align}
The $a_f$ and $b_f$ are the SM vector and
axial-vector couplings. 
The most general $X_1WW$ interaction at the lowest dimension can be
written as~\cite{Hagiwara:1986vm}  
\begin{align}
 {\cal L}_1^W =
    &\,i\kappa_{\sss W_1}g_{\sss WWZ} (W^+_{\mu\nu}W^{- \mu} - W^-_{\mu\nu}W^{+\mu})
 X_{1}^{\nu} 
   + i\kappa_{\sss W_2}g_{\sss WWZ} W^+_\mu W^-_\nu X_{1}^{\mu\nu}  
  -\kappa_{\sss W_3} W^+_\mu W^-_\nu(\partial^\mu X_{1}^{\nu} +
 \partial^\nu X_{1}^{\mu})\nn\\
  &+ i\kappa_{\sss W_4} W^+_\mu W^-_\nu \widetilde X_{1}^{\mu\nu}
  - \kappa_{\sss W_5} \epsilon_{\mu\nu\rho\sigma}
  [{W^+}^\mu ({\partial}^\rho {W^-}^\nu)-({\partial}^\rho {W^+}^\mu){W^-}^\nu] X_{1}^{\sigma}\,,
\end{align}
where $g_{WWZ}=-e\cot\theta_W$. 
Note that our effective field
theory description lives at energy scales where EW symmetry $SU(2)_L \times
U(1)_Y$ is broken to $U(1)_{EM}$. This approach does not require to specify
the transformation properties of $X_1$ with respect to the EW symmetry.  
In the case of $ZZ$, 
Bose symmetry implies a reduction of the possible terms and the interaction Lagrangian reduces to~\cite{Hagiwara:1986vm,Keung:2008ve}
\begin{align}
 {\cal L}_1^Z =
  -\kappa_{\sss Z_1} Z_{\mu\nu}Z^{\mu} X_{1}^{\nu}
  -\kappa_{\sss Z_3} X_{1}^{\mu}({\partial}^{\nu} Z_{\mu})Z_{\nu} 
  -\kappa_{\sss Z_5} \epsilon_{\mu\nu\rho\sigma}  X_{1}^{\mu}
  Z^{\nu} ({\partial}^\rho Z^{\sigma})\,. 
\label{eq:1zz}
\end{align}
Parity
conservation implies that 
$\kappa_{\sss f_b}= \kappa_{\sss V_4}=\kappa_{\sss V_5}= 0$ 
for $X_1=1^-$ while 
$\kappa_{\sss f_a}=\kappa_{\sss V_1}=\kappa_{\sss V_2}=\kappa_{\sss V_3} = 0$
for $X_1=1^+$.

\subsection{Spin 2}

The interaction lagrangian for the spin-2 boson proceeds via the
energy-momentum (E-M) tensor of the SM fields and starts at
dimension five~\cite{Giudice:1998ck,Han:1998sg}:
\begin{align}
 {\cal L}_2^f = -\frac{1}{\Lambda}\sum_{f=q,\ell} 
  \kappa_{f}\,T^f_{\mu\nu}X_2^{\mu\nu} \quad {\rm and}\quad
 {\cal L}_2^V = -\frac{1}{\Lambda}\sum_{V=Z,W,\gamma,g} 
  \kappa_{V}\,T^V_{\mu\nu}X_2^{\mu\nu},
\end{align}
where $T^{f,V}_{\mu\nu}$ are the E-M tensors; see
refs.~\cite{Han:1998sg,Hagiwara:2008jb} for the explicit forms.
The coupling parameters $\kappa_f$ and $\kappa_V$ are 
introduced~\cite{Ellis:2012jv,Englert:2012xt} in full analogy with what
has been done in the spin-0 and -1 cases.
For $X_2=2^+$ in the minimal RS-like graviton scenario, i.e. the universal
coupling strength to the matter and gauge fields, the parameters should
be chosen as $\kappa_{f}=\kappa_{V}\neq 0$.

\section{Distributions in the $X\to 4\ell$ analysis: Comparison with the
 JHU program}

\begin{figure}
\centering
\quad spin-0 \hspace*{3.75cm} spin-1 \hspace*{3.75cm} spin-2 
 \\
 \includegraphics[width=0.275\textwidth]{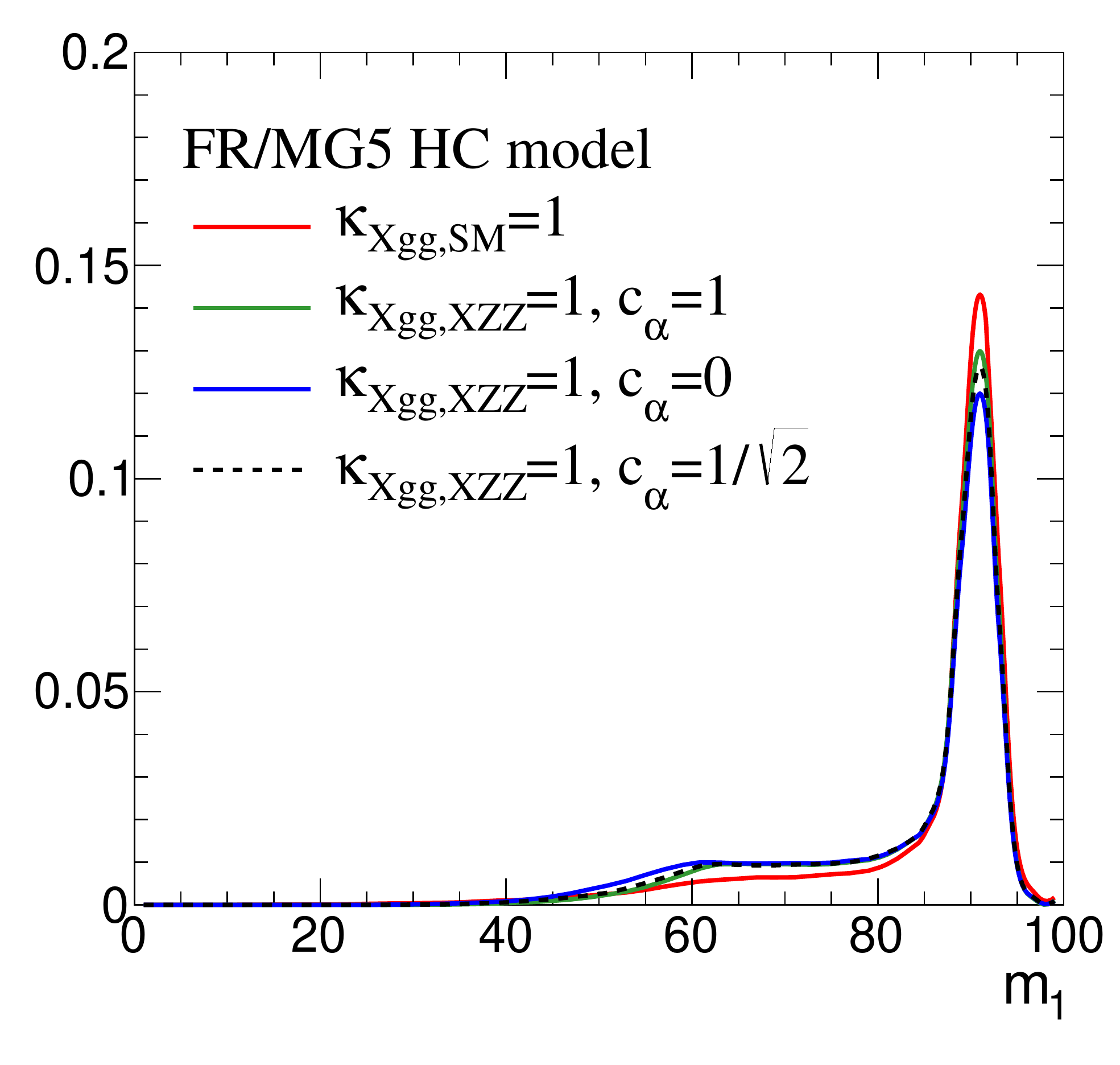}
 \includegraphics[width=0.275\textwidth]{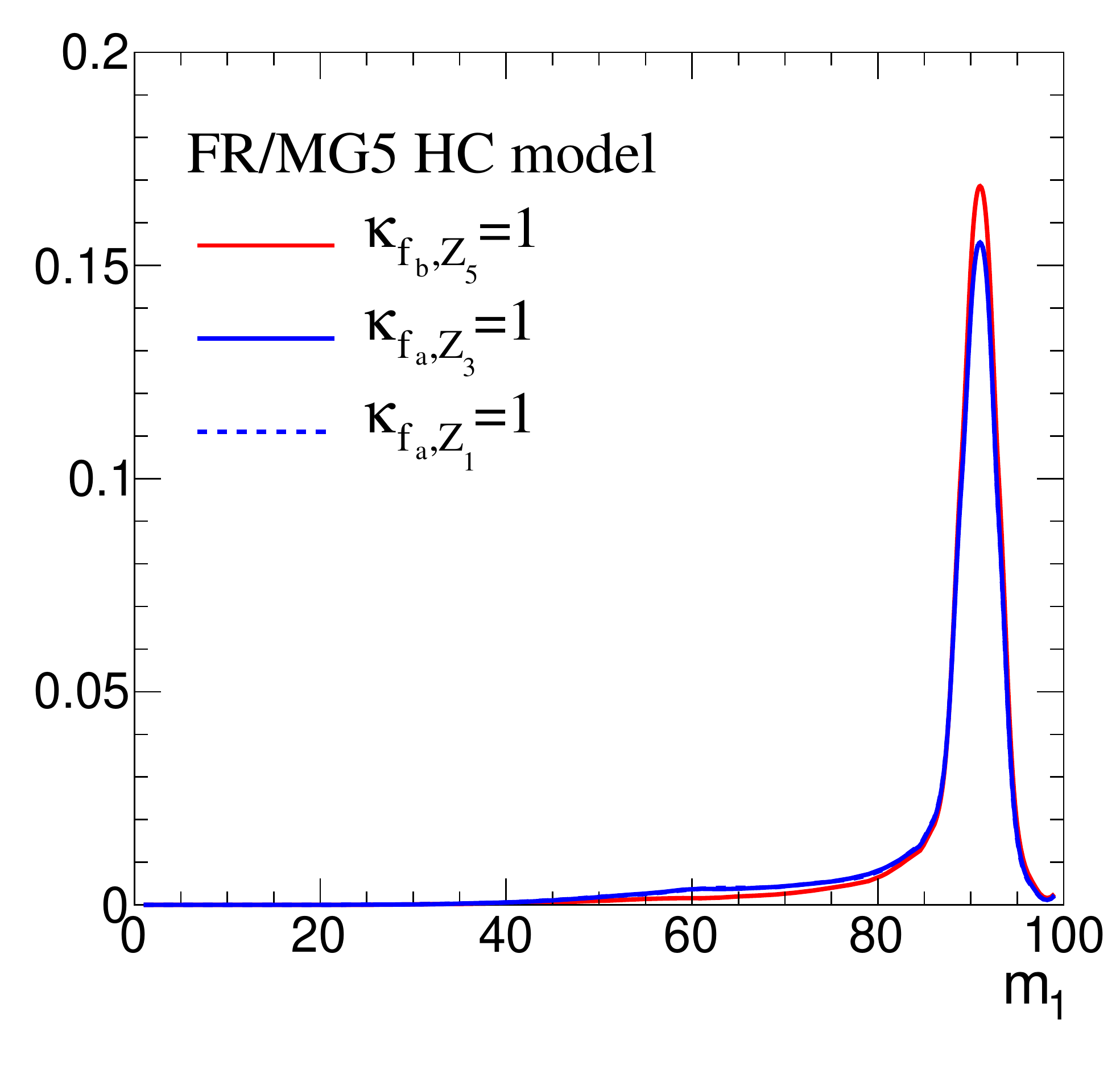}
 \includegraphics[width=0.275\textwidth]{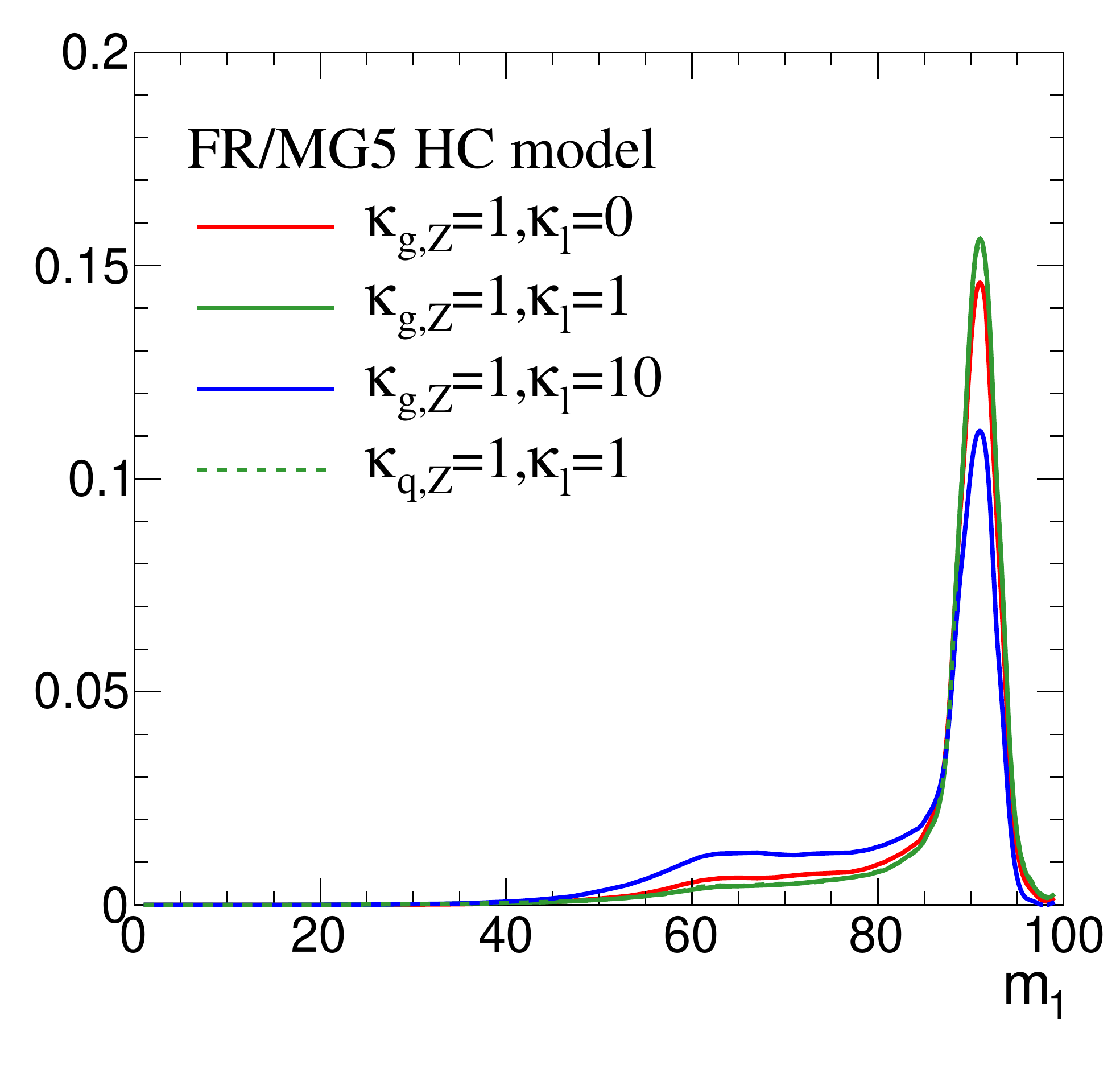} \\
 \includegraphics[width=0.275\textwidth]{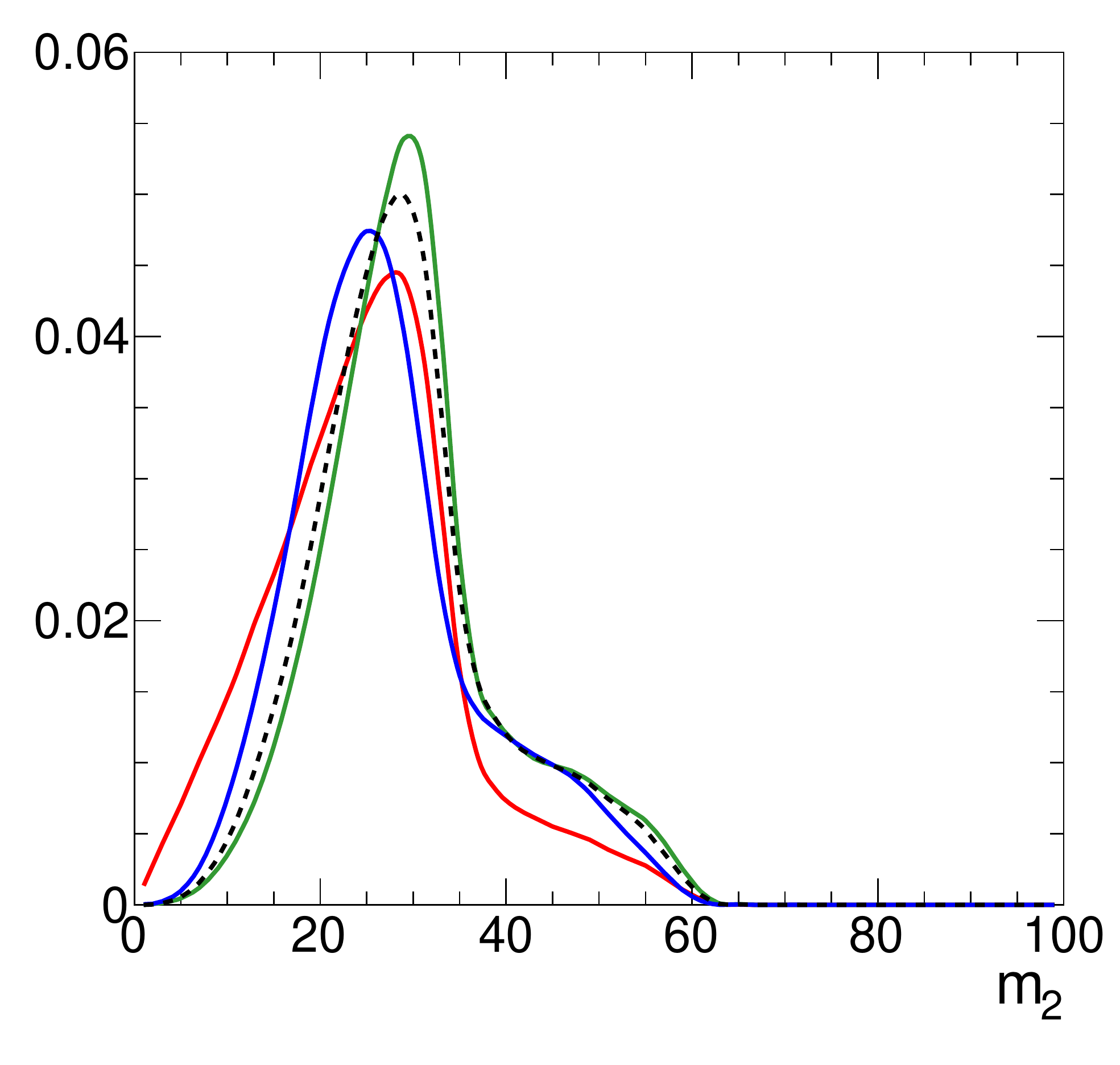}
 \includegraphics[width=0.275\textwidth]{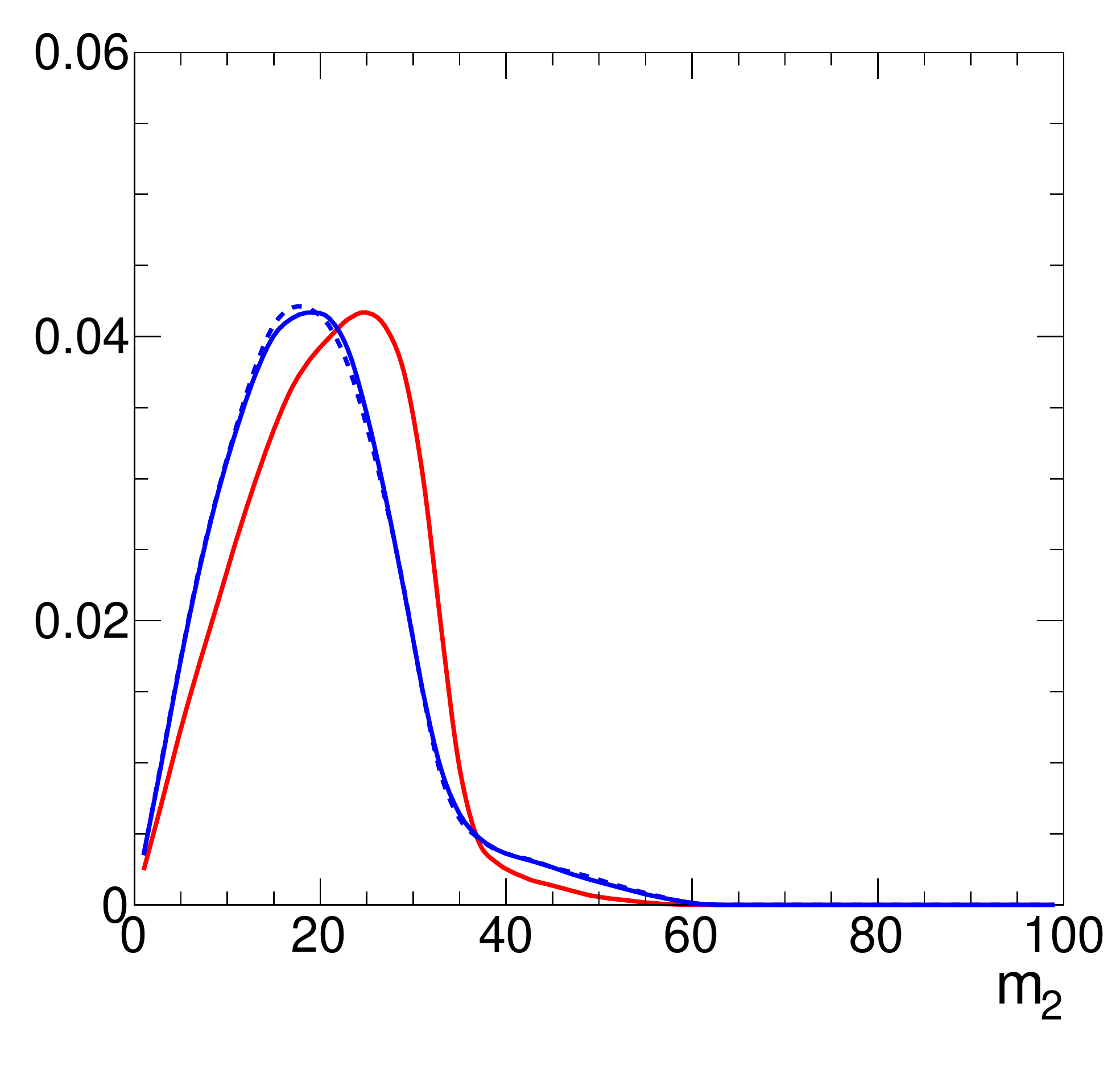}
 \includegraphics[width=0.275\textwidth]{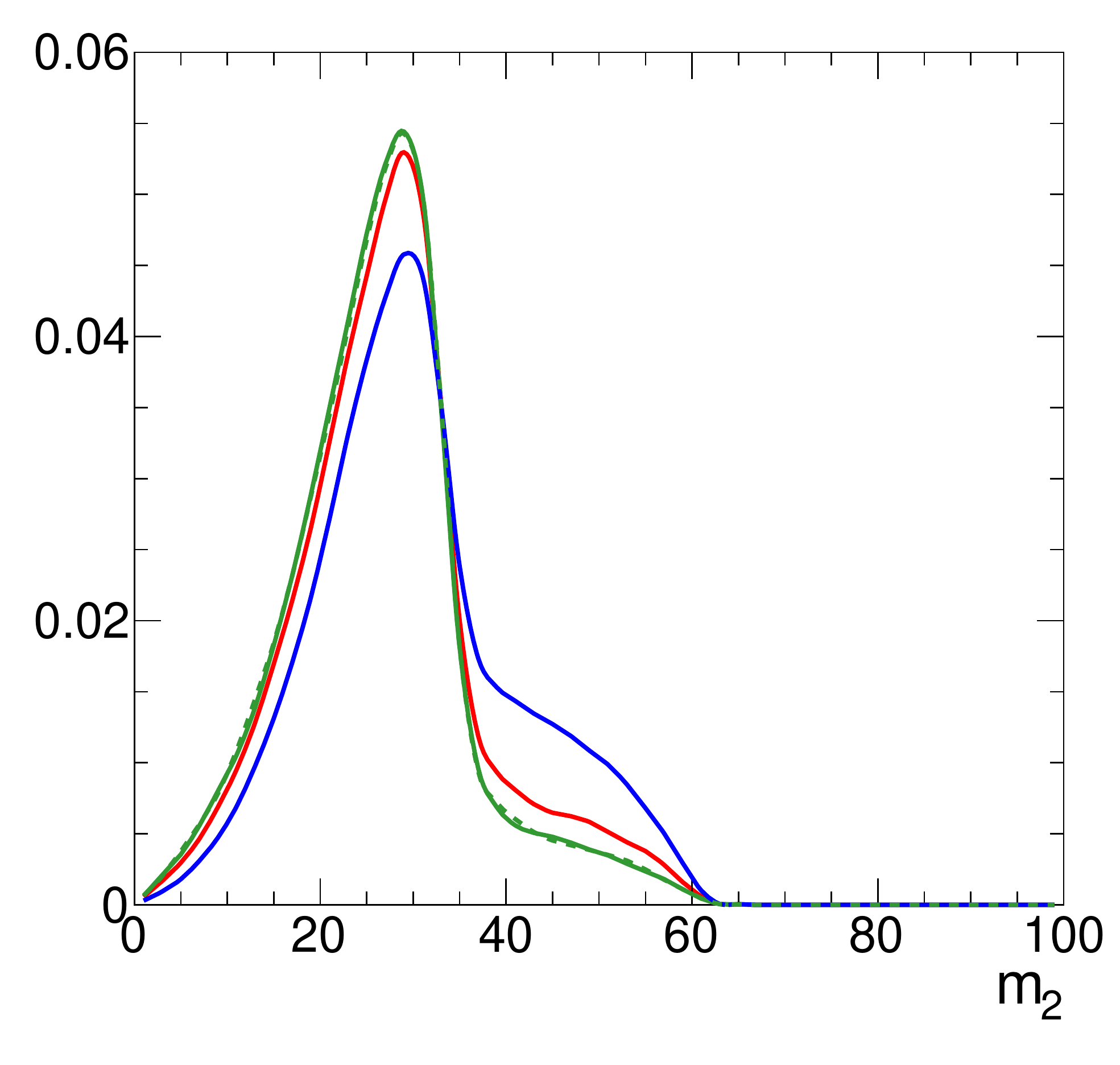}
\caption{Normalised distributions of the lepton invariant masses
 in the $X\to ZZ$ analysis
 (cf. fig.~11 in the JHU paper~\cite{Bolognesi:2012mm}).}
\label{fig:ZZ_m}
\end{figure}

\begin{figure}
\centering
\quad spin-0 \hspace*{3.75cm} spin-1 \hspace*{3.75cm} spin-2 
 \\ 
 \includegraphics[width=0.275\textwidth]{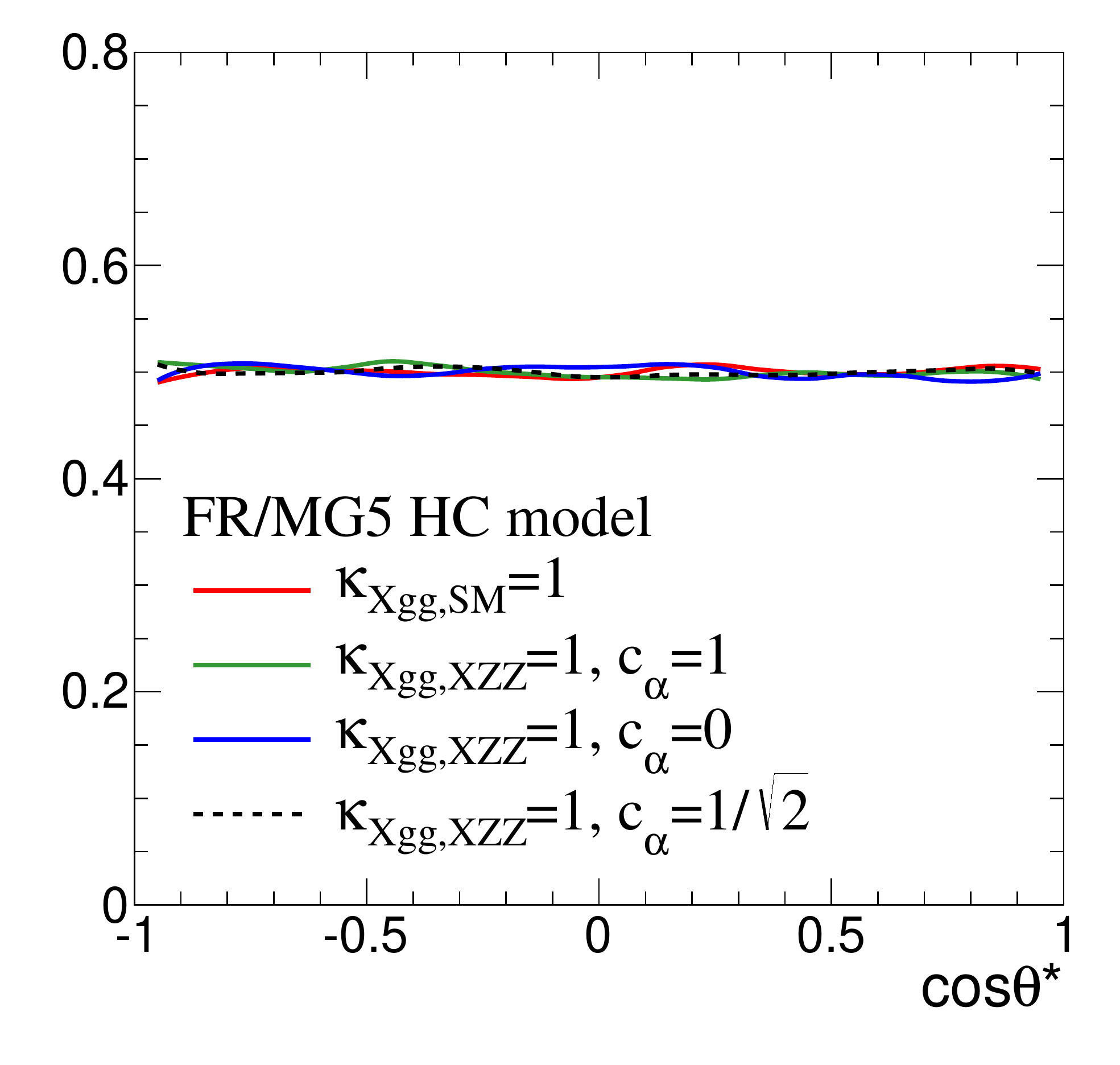}
 \includegraphics[width=0.275\textwidth]{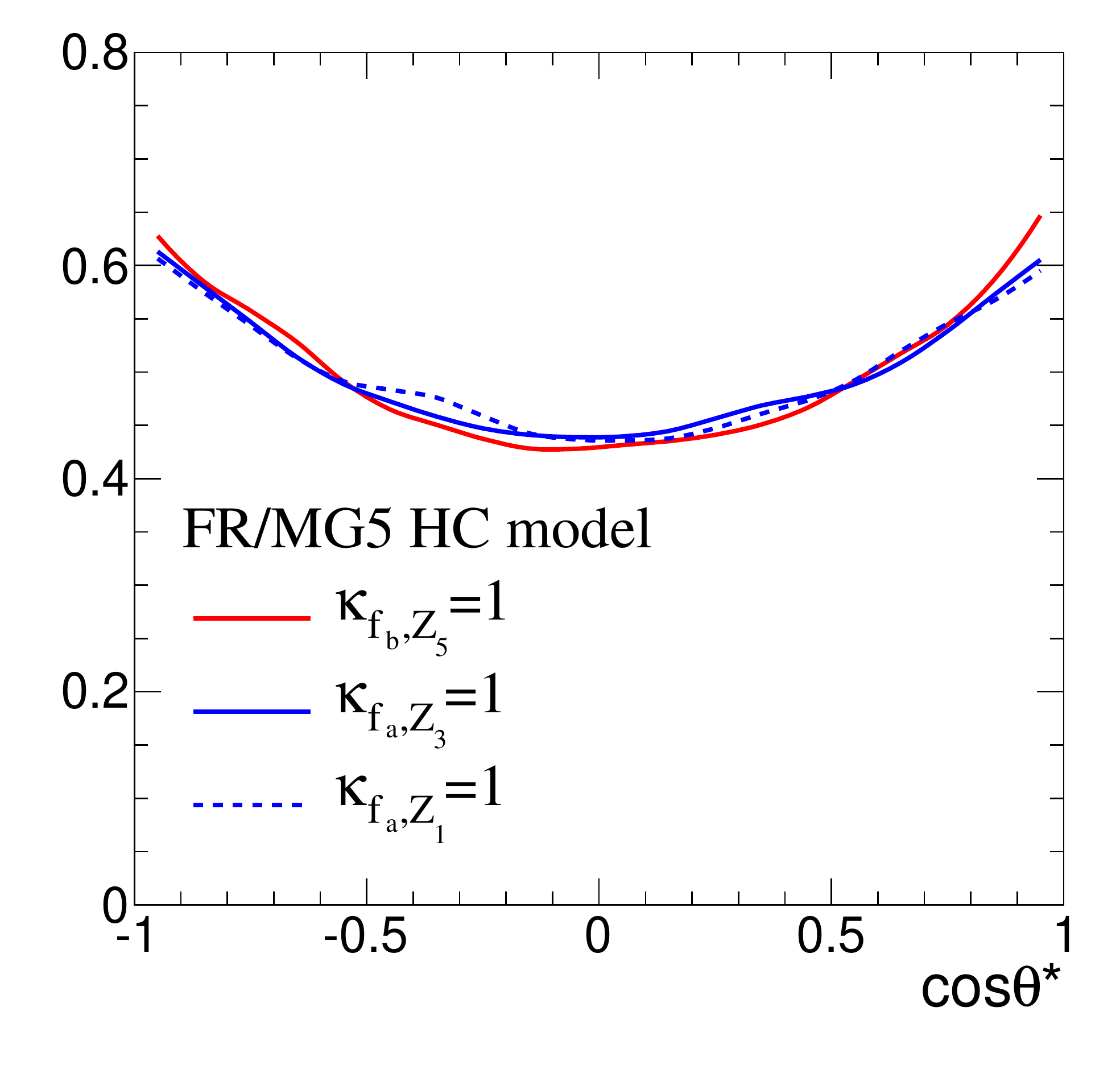}
 \includegraphics[width=0.275\textwidth]{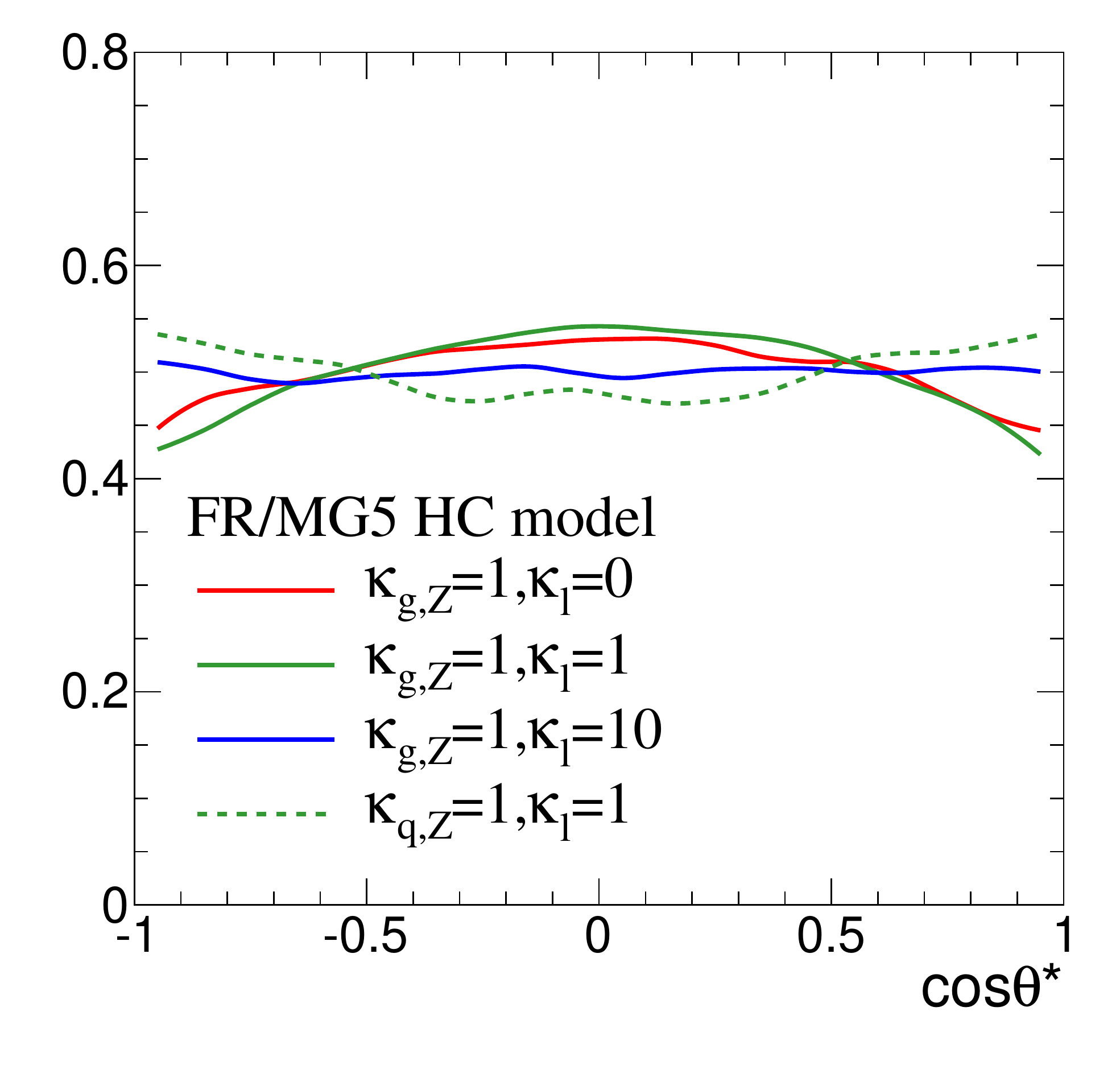} \\
 \includegraphics[width=0.275\textwidth]{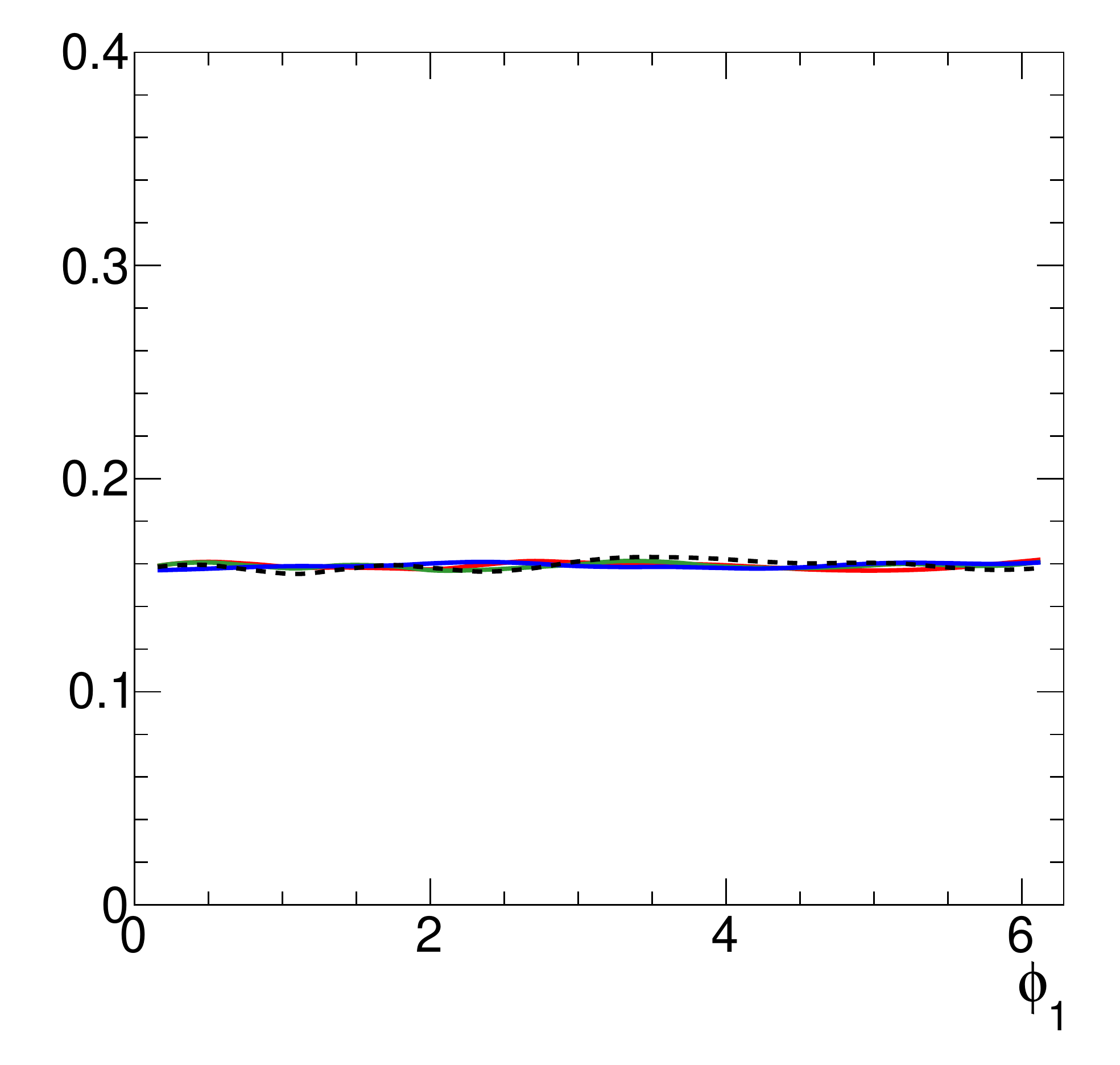}
 \includegraphics[width=0.275\textwidth]{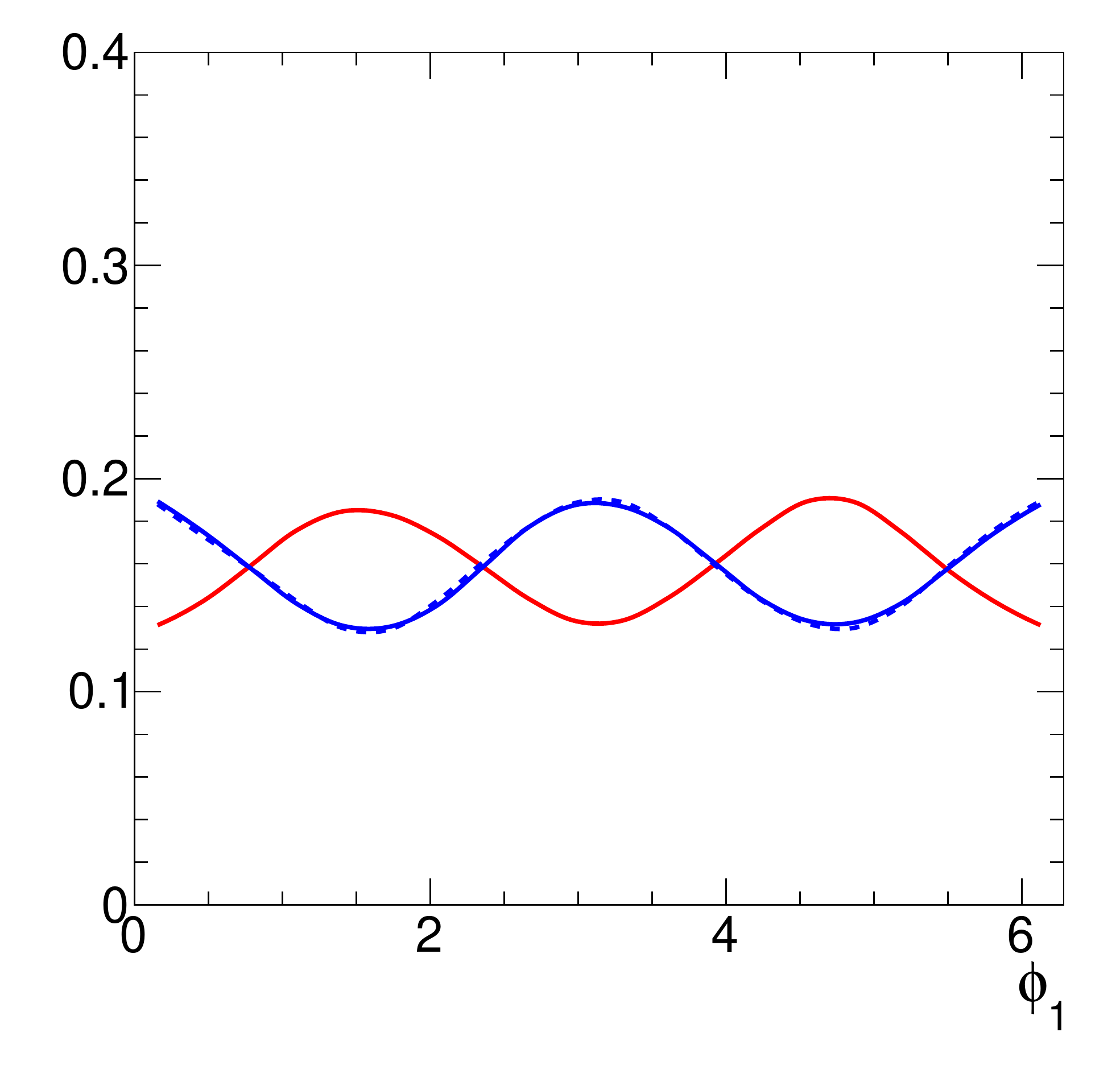}
 \includegraphics[width=0.275\textwidth]{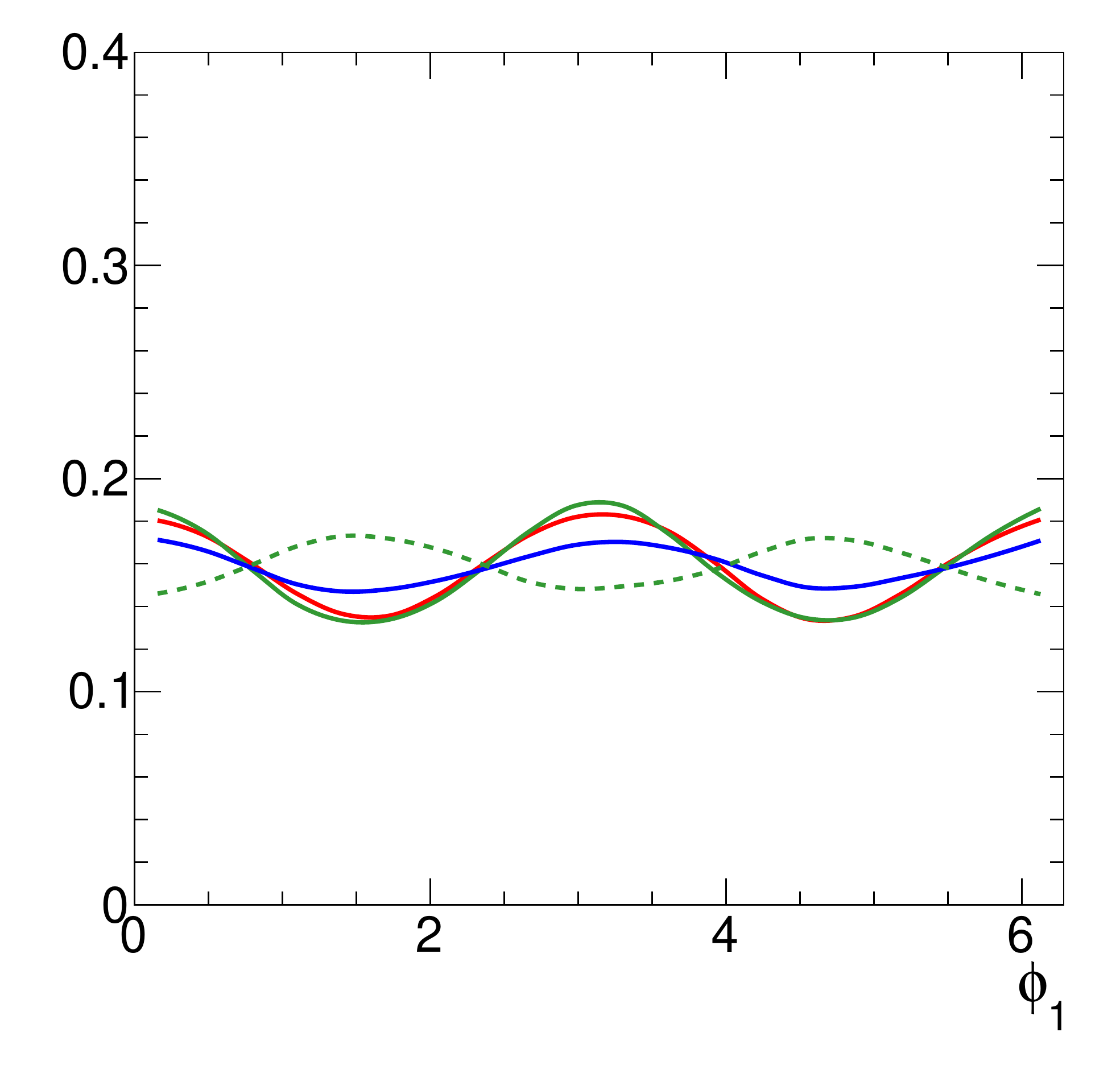} \\
 \includegraphics[width=0.275\textwidth]{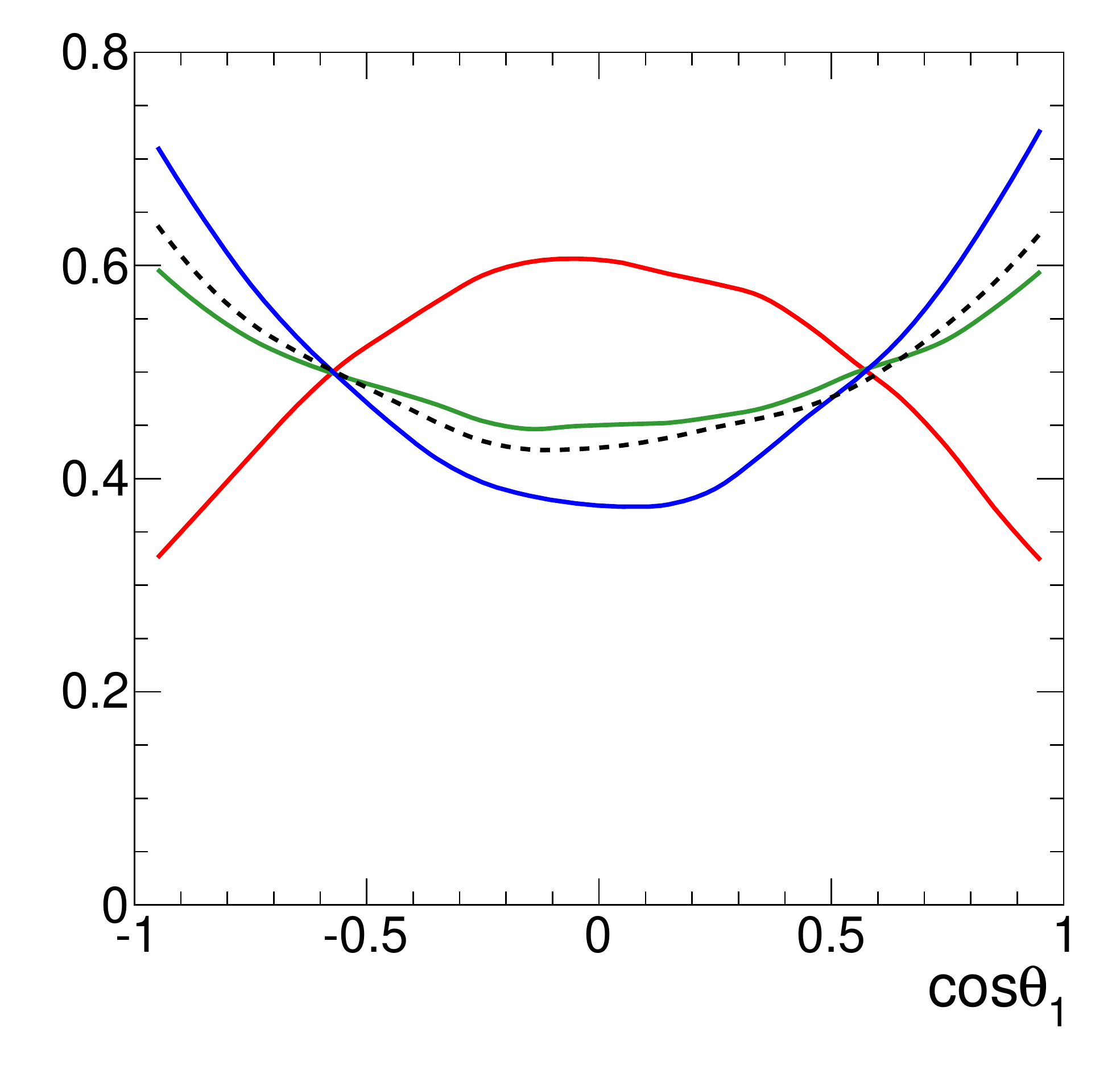}
 \includegraphics[width=0.275\textwidth]{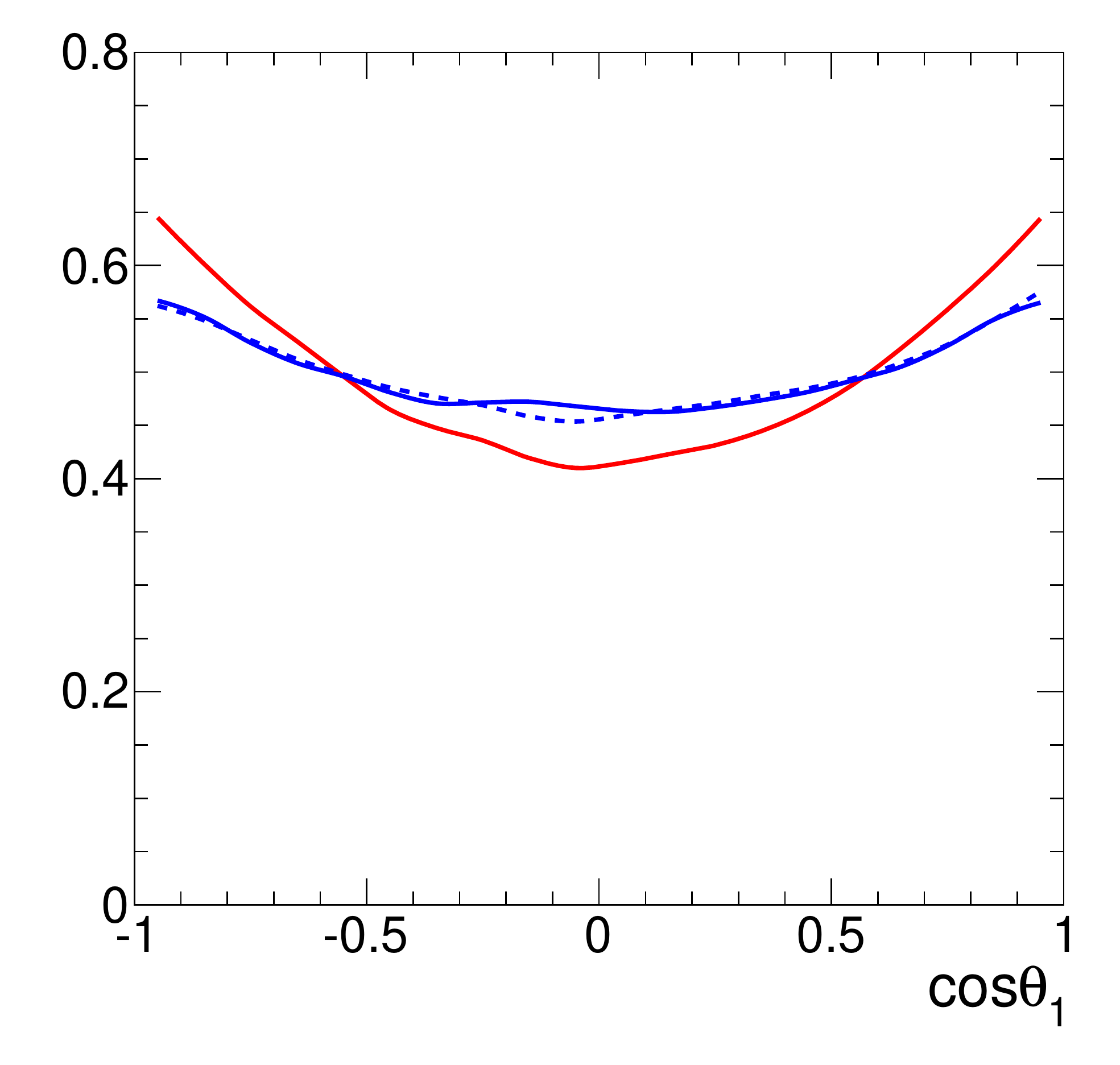}
 \includegraphics[width=0.275\textwidth]{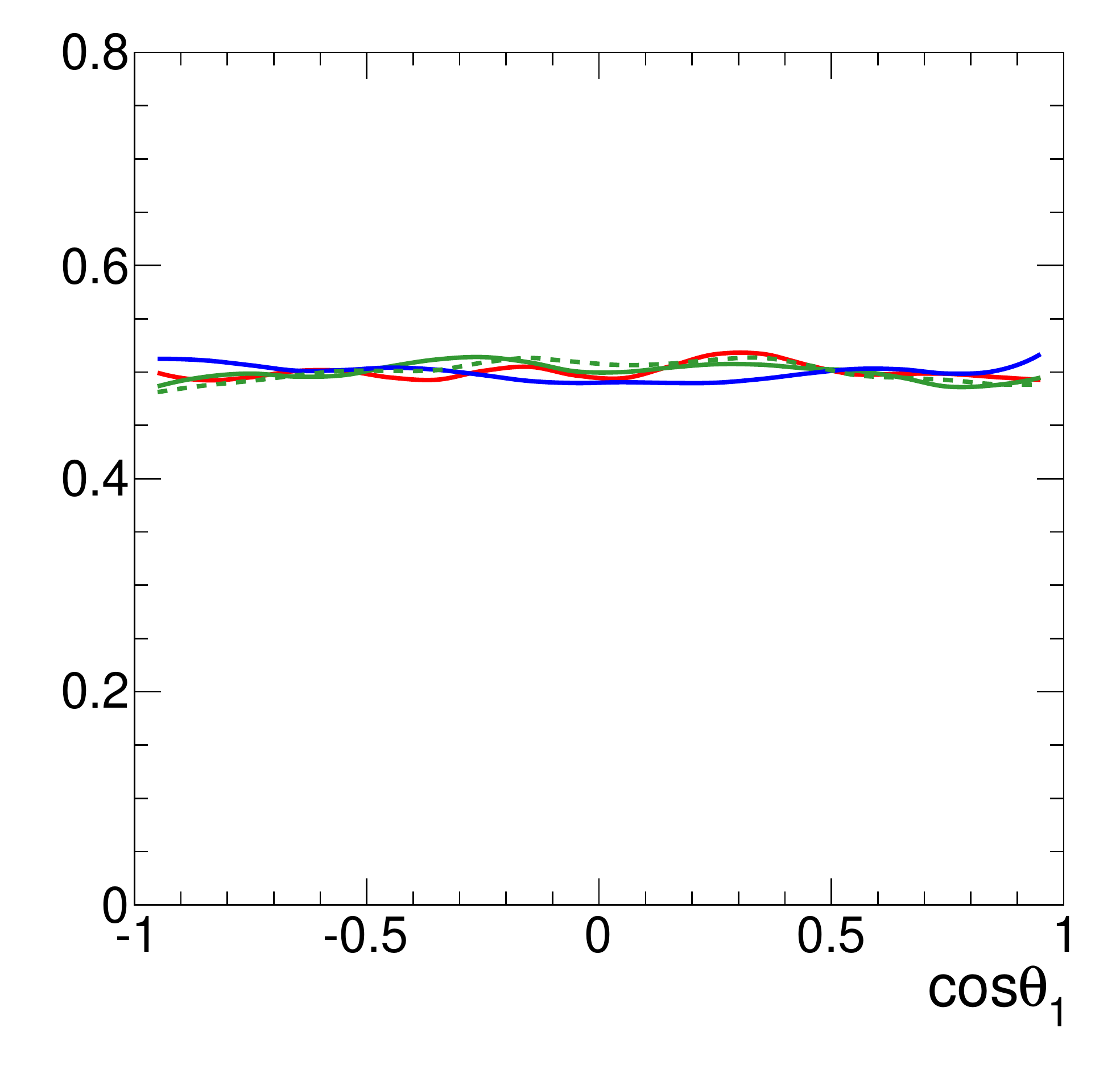} \\
 \includegraphics[width=0.275\textwidth]{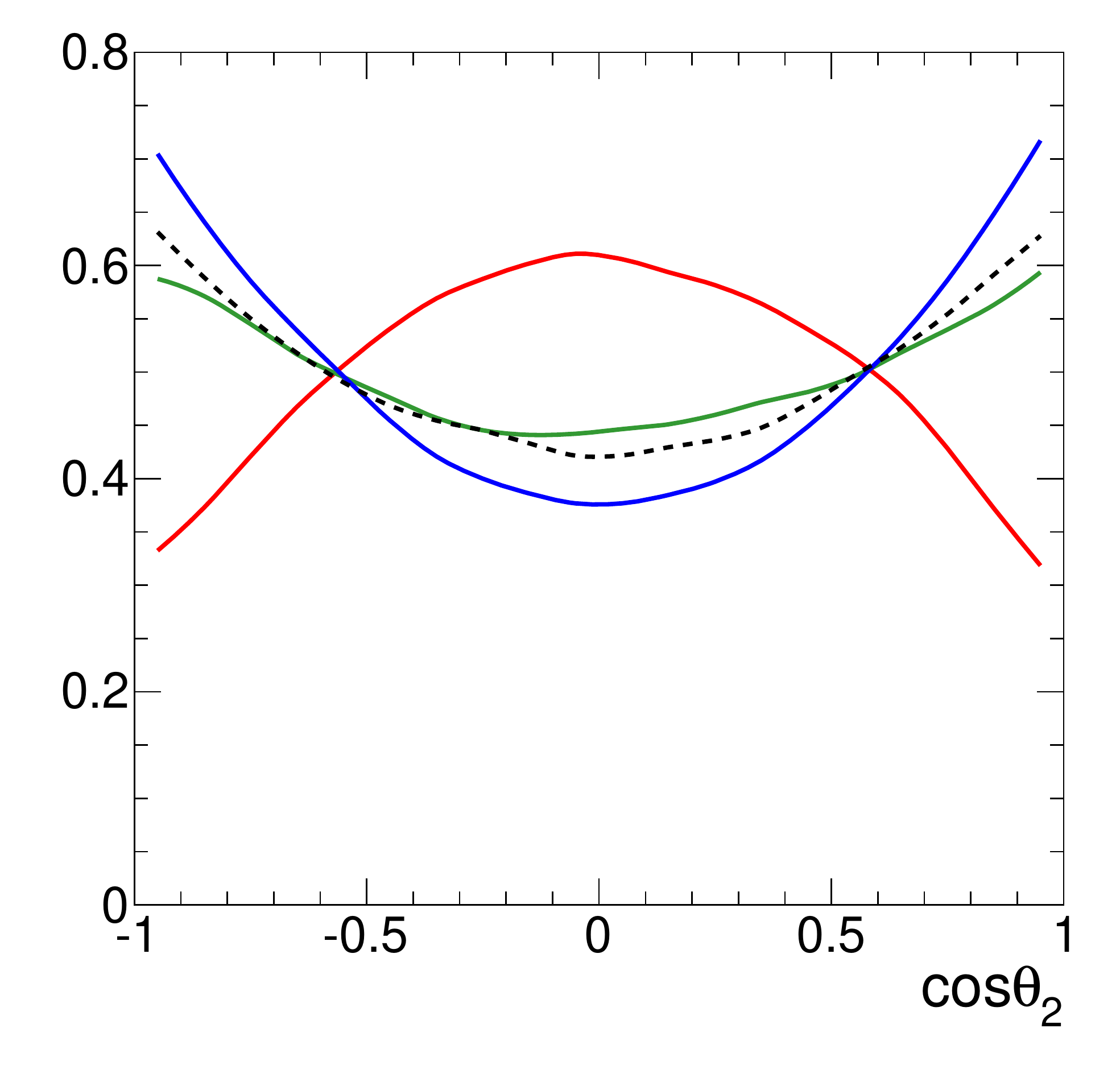}
 \includegraphics[width=0.275\textwidth]{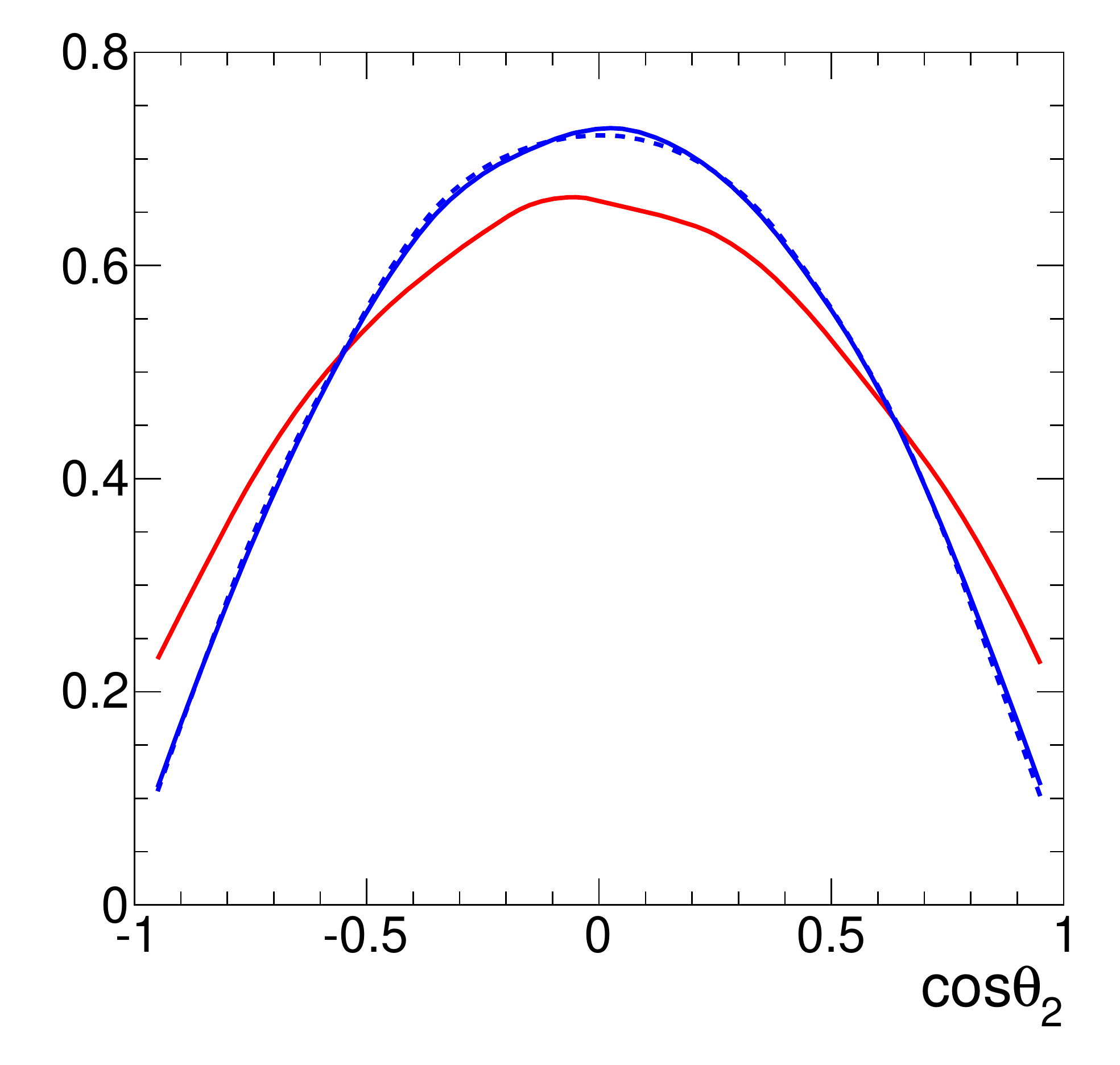}
 \includegraphics[width=0.275\textwidth]{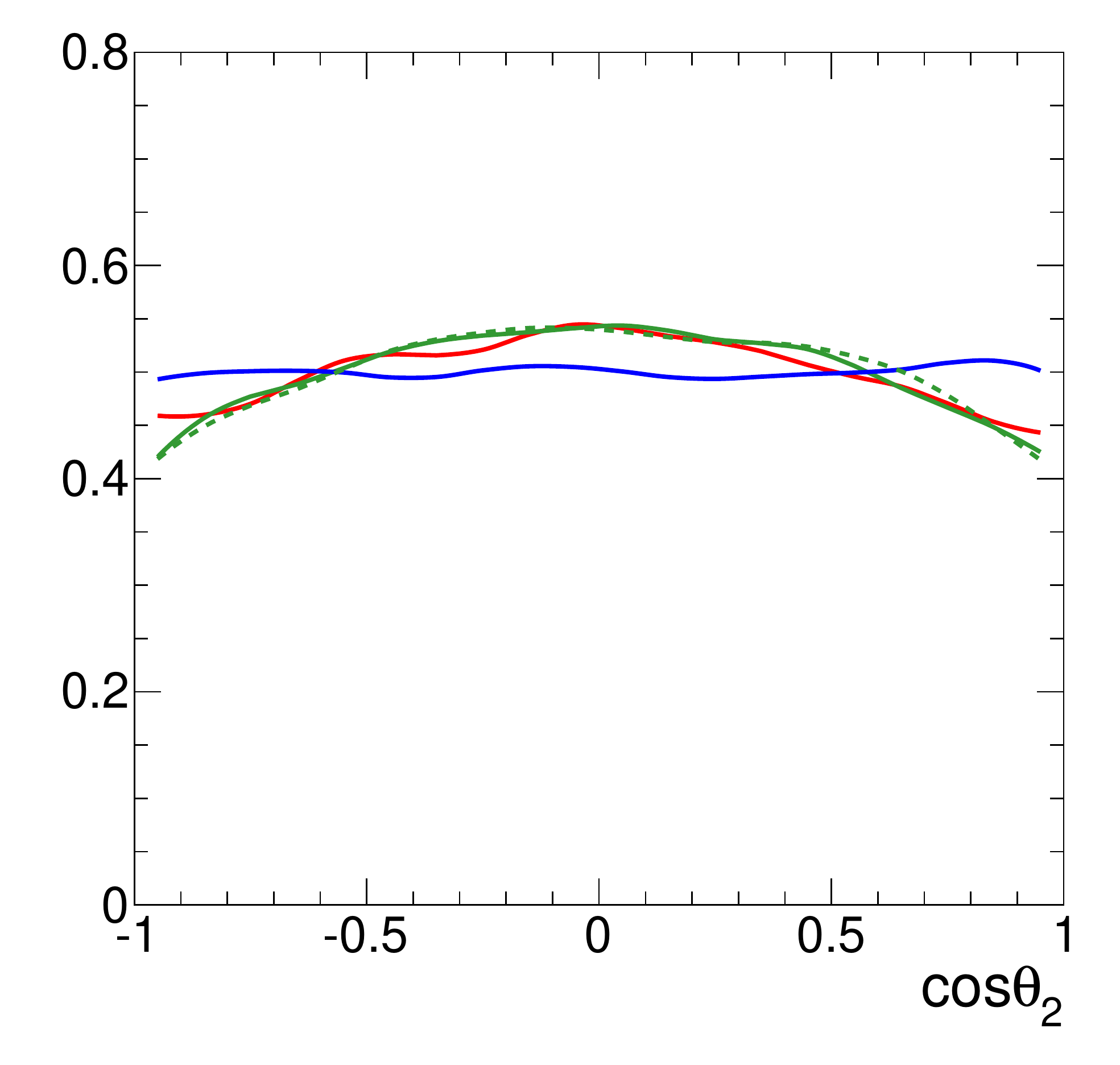} \\
 \includegraphics[width=0.275\textwidth]{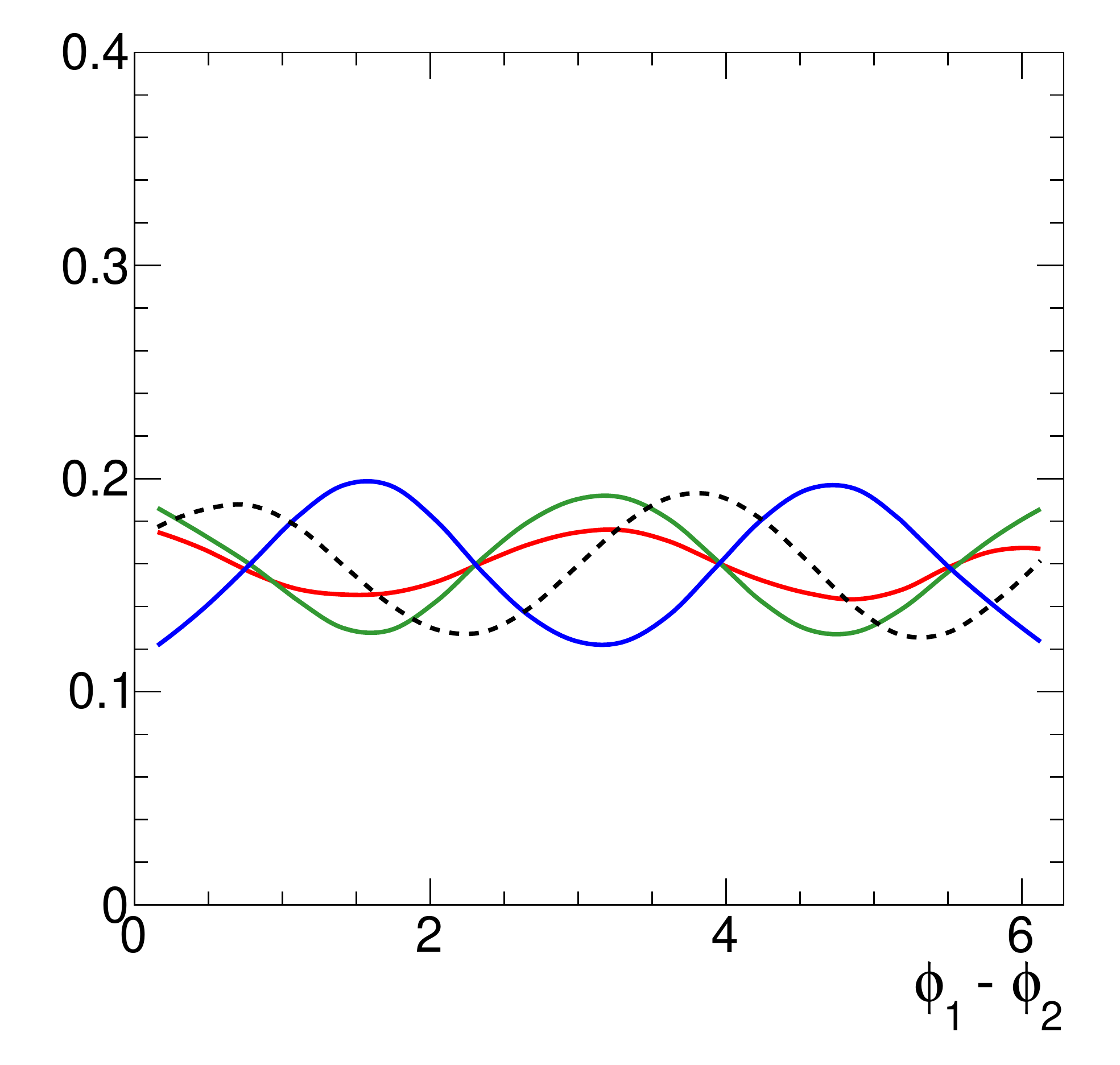}
 \includegraphics[width=0.275\textwidth]{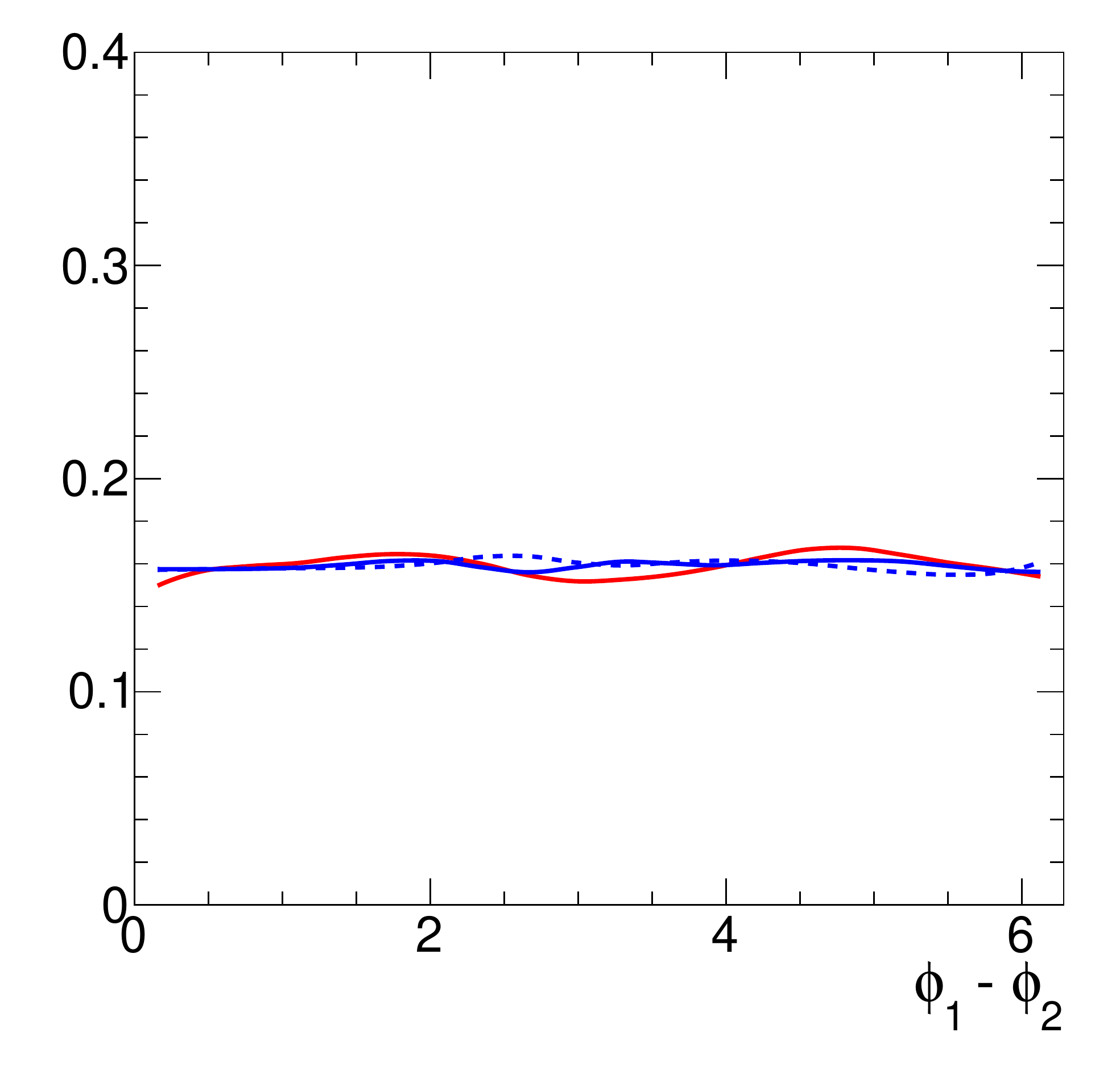}
 \includegraphics[width=0.275\textwidth]{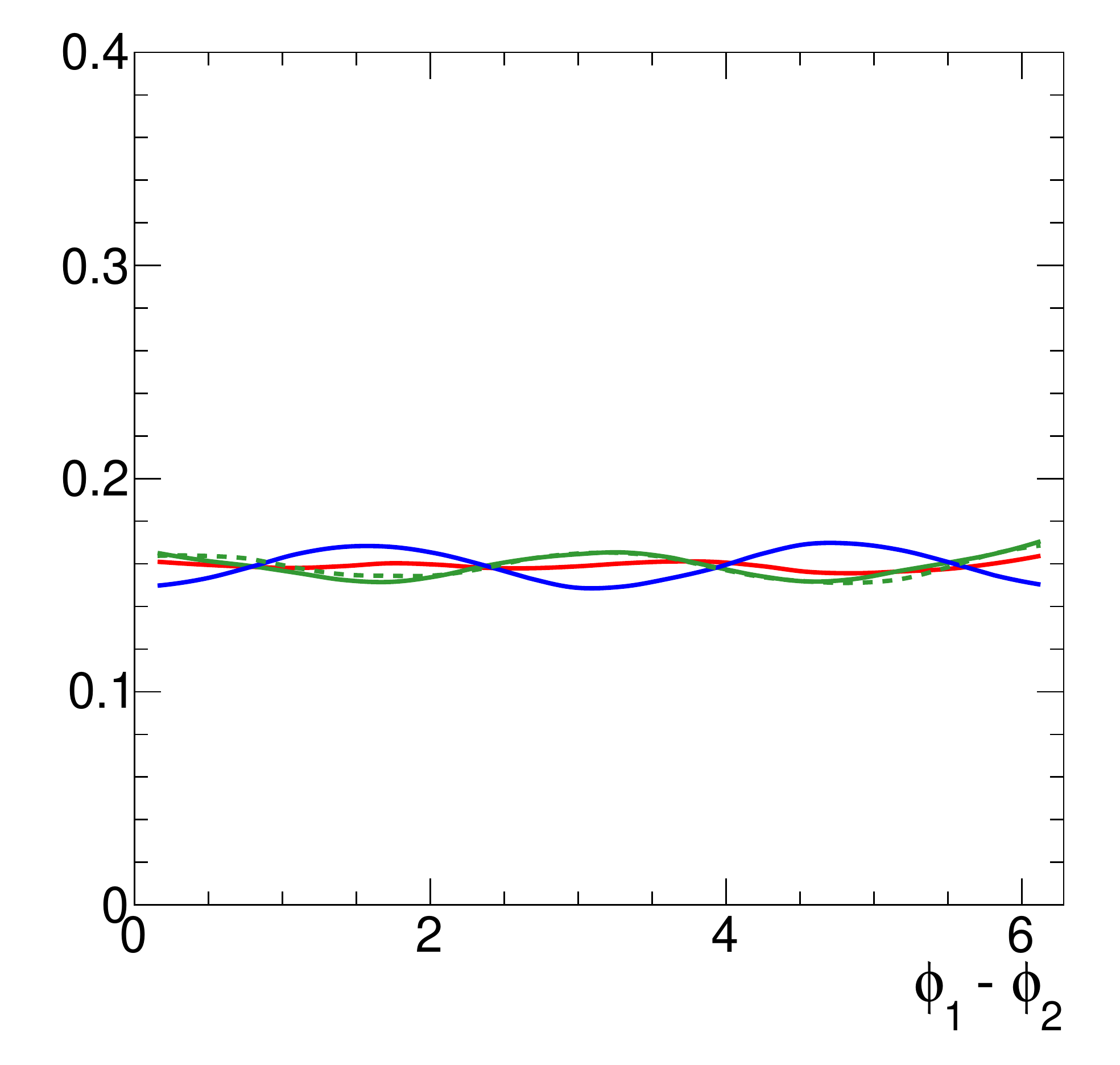} 
\caption{Angular distributions in the $X\to ZZ$ analysis
 (cf. fig.~12 in the JHU paper~\cite{Bolognesi:2012mm}).}
\label{fig:ZZ}
\end{figure}

\begin{figure}
\quad spin-0 \hspace*{3.75cm} spin-1 \hspace*{3.75cm} spin-2 
 \\
 \includegraphics[width=0.275\textwidth]{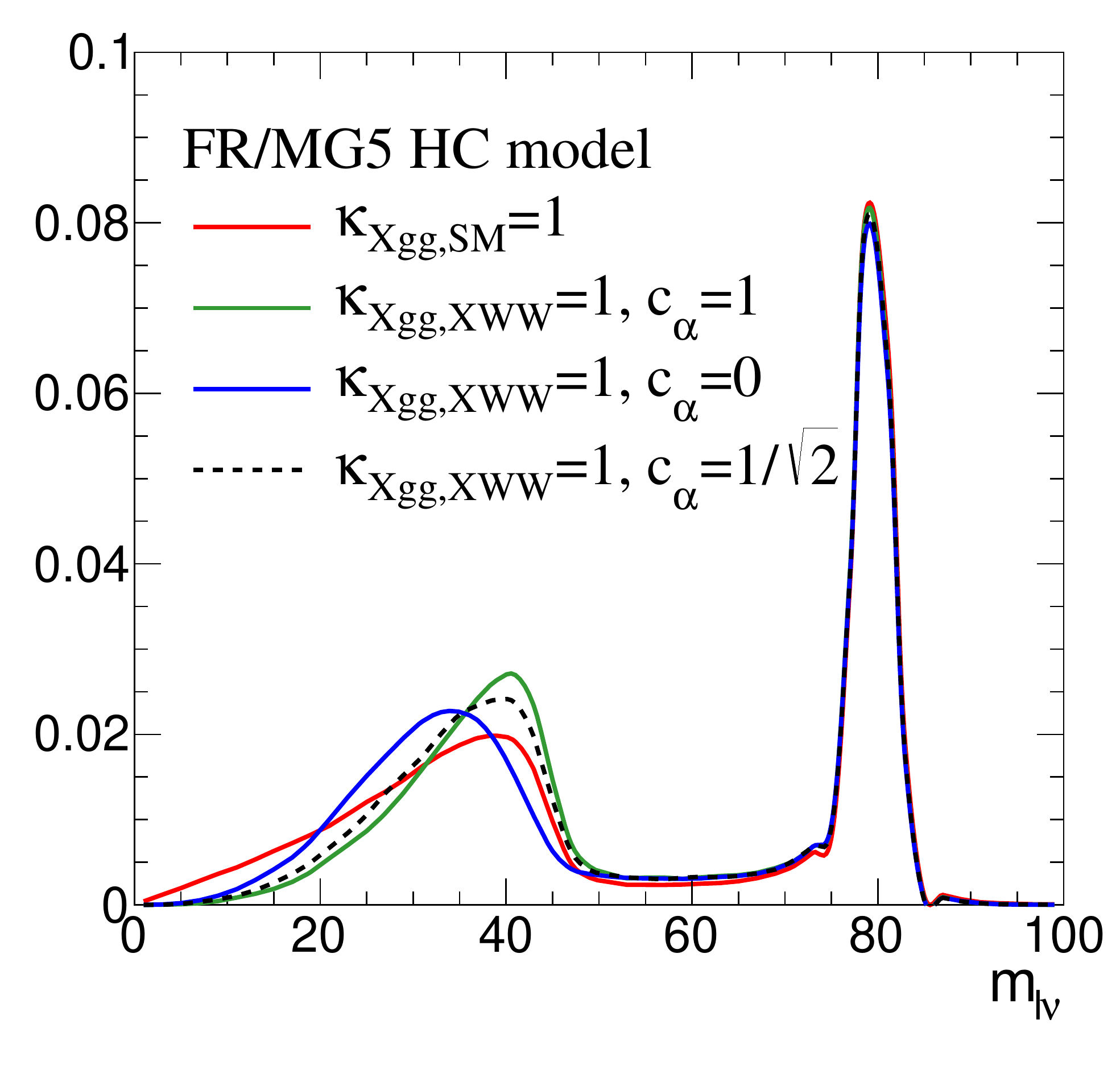}
 \includegraphics[width=0.275\textwidth]{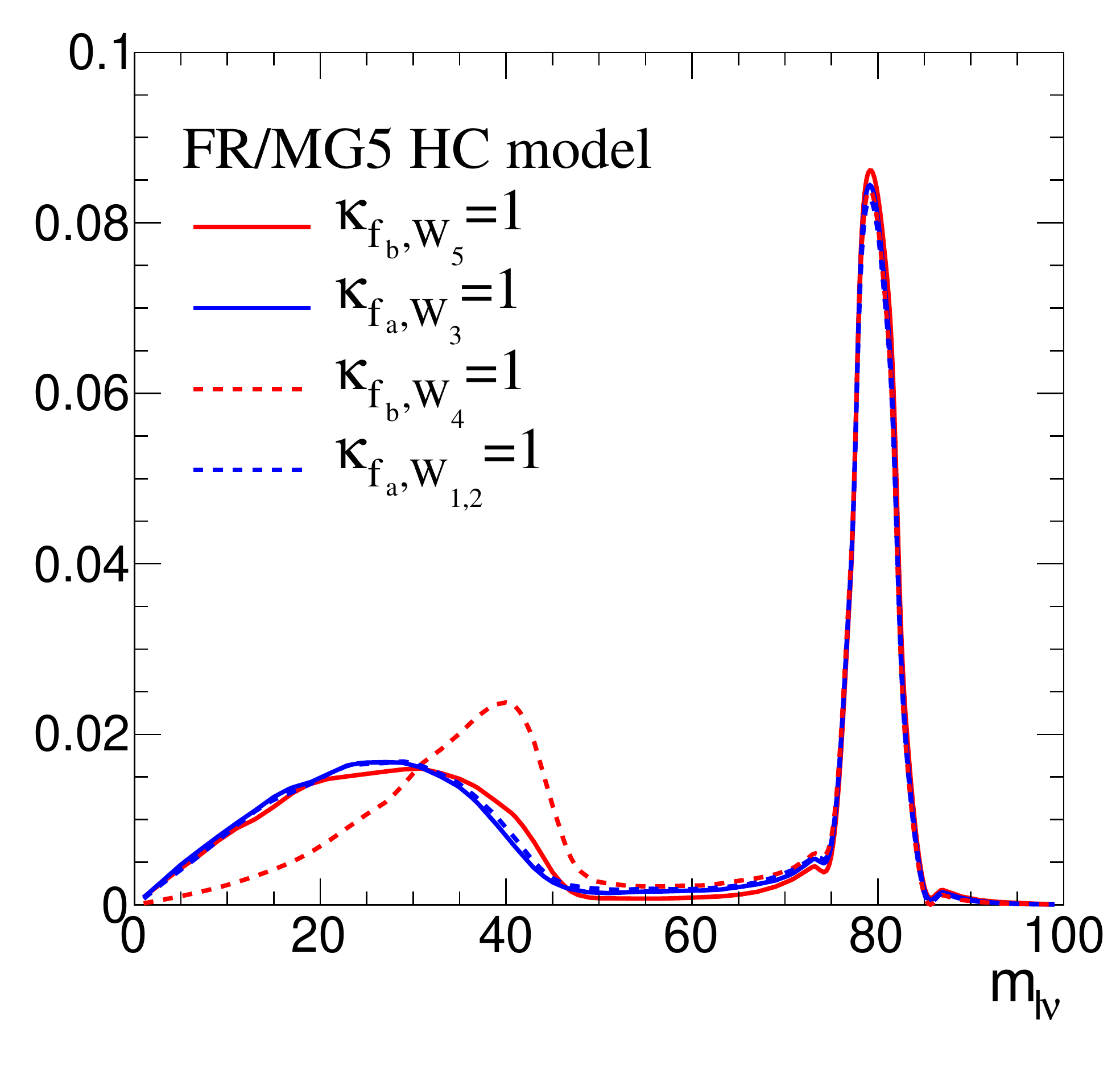}
 \includegraphics[width=0.275\textwidth]{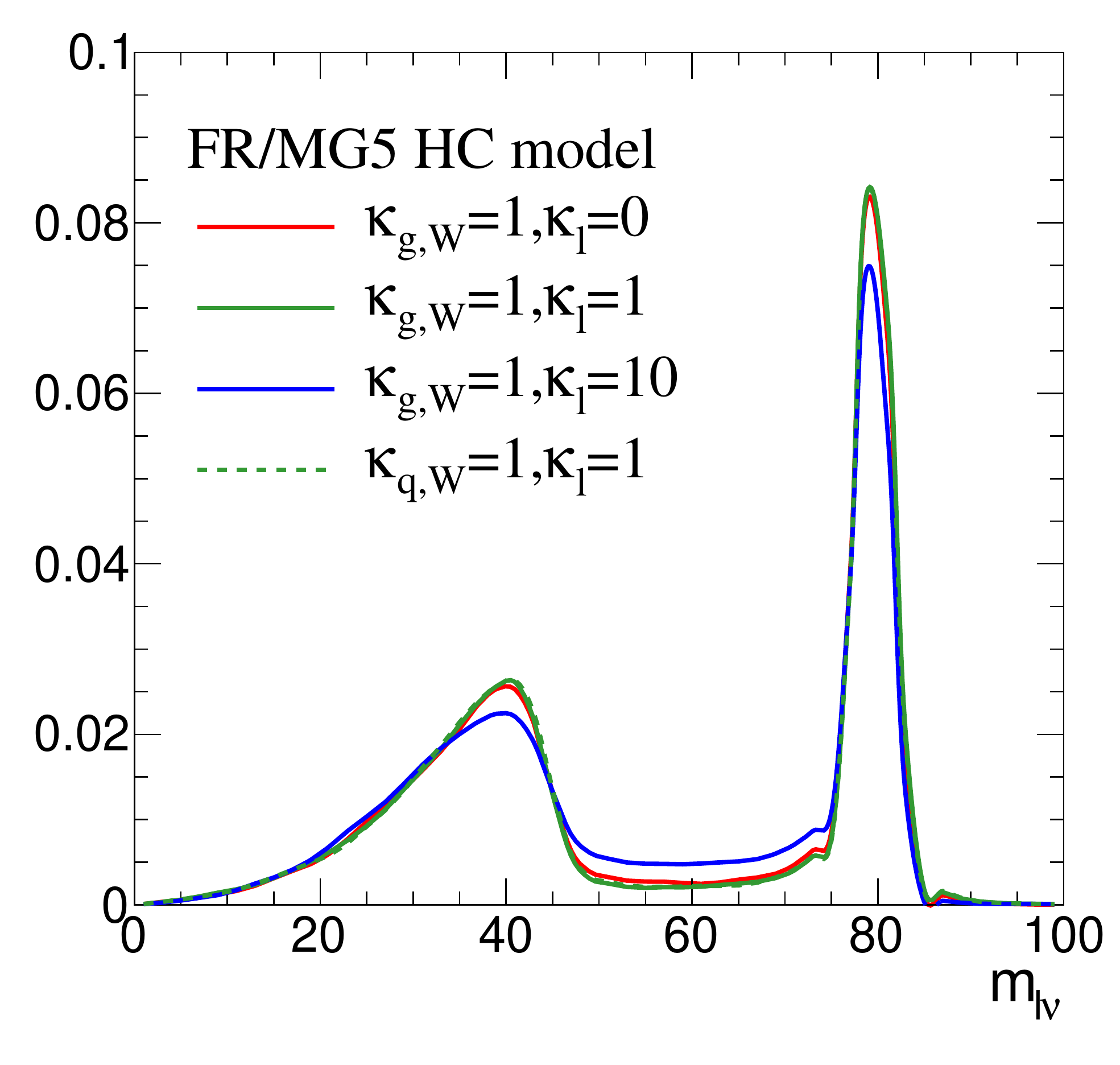} \\
 \includegraphics[width=0.275\textwidth]{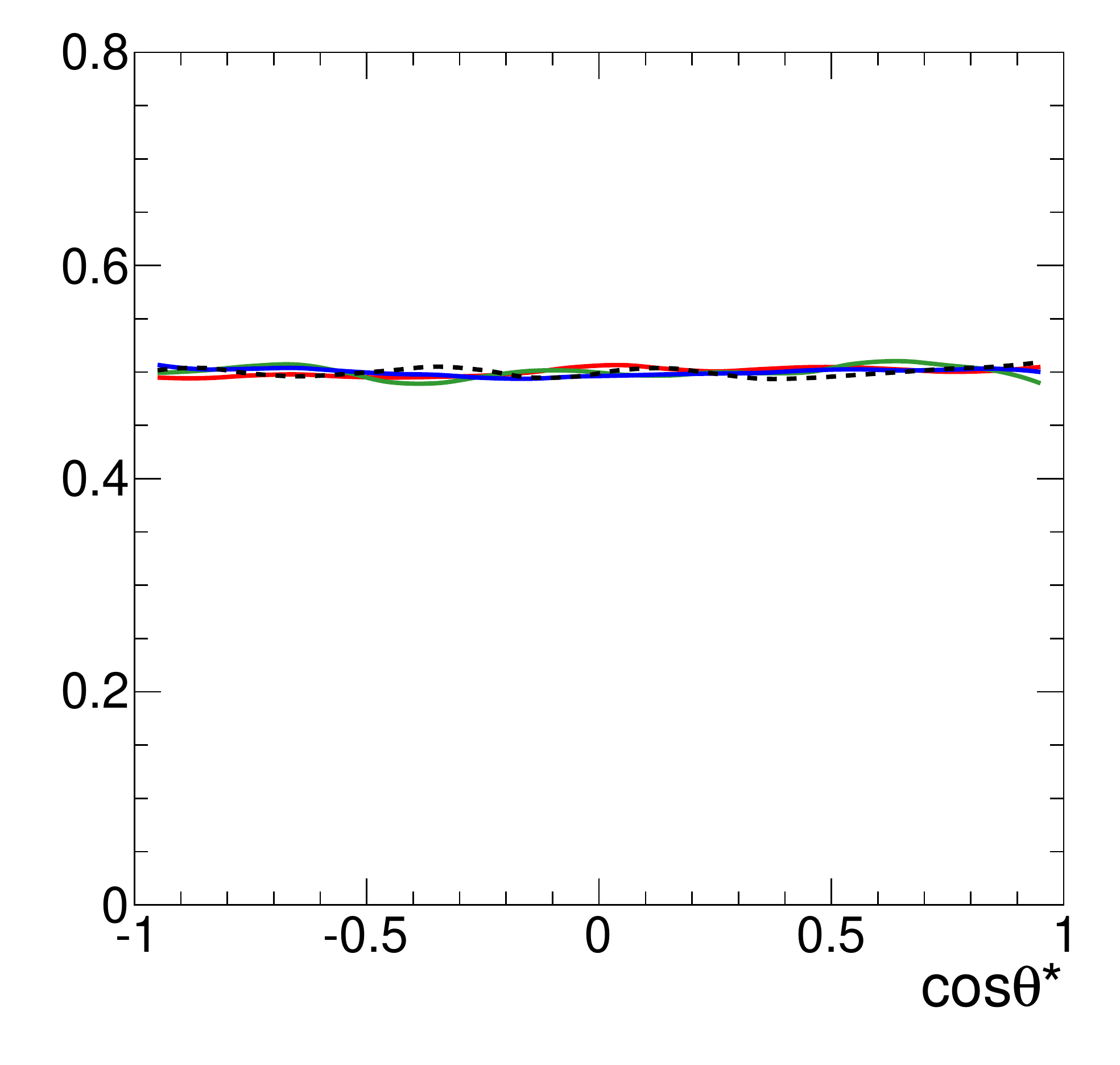}
 \includegraphics[width=0.275\textwidth]{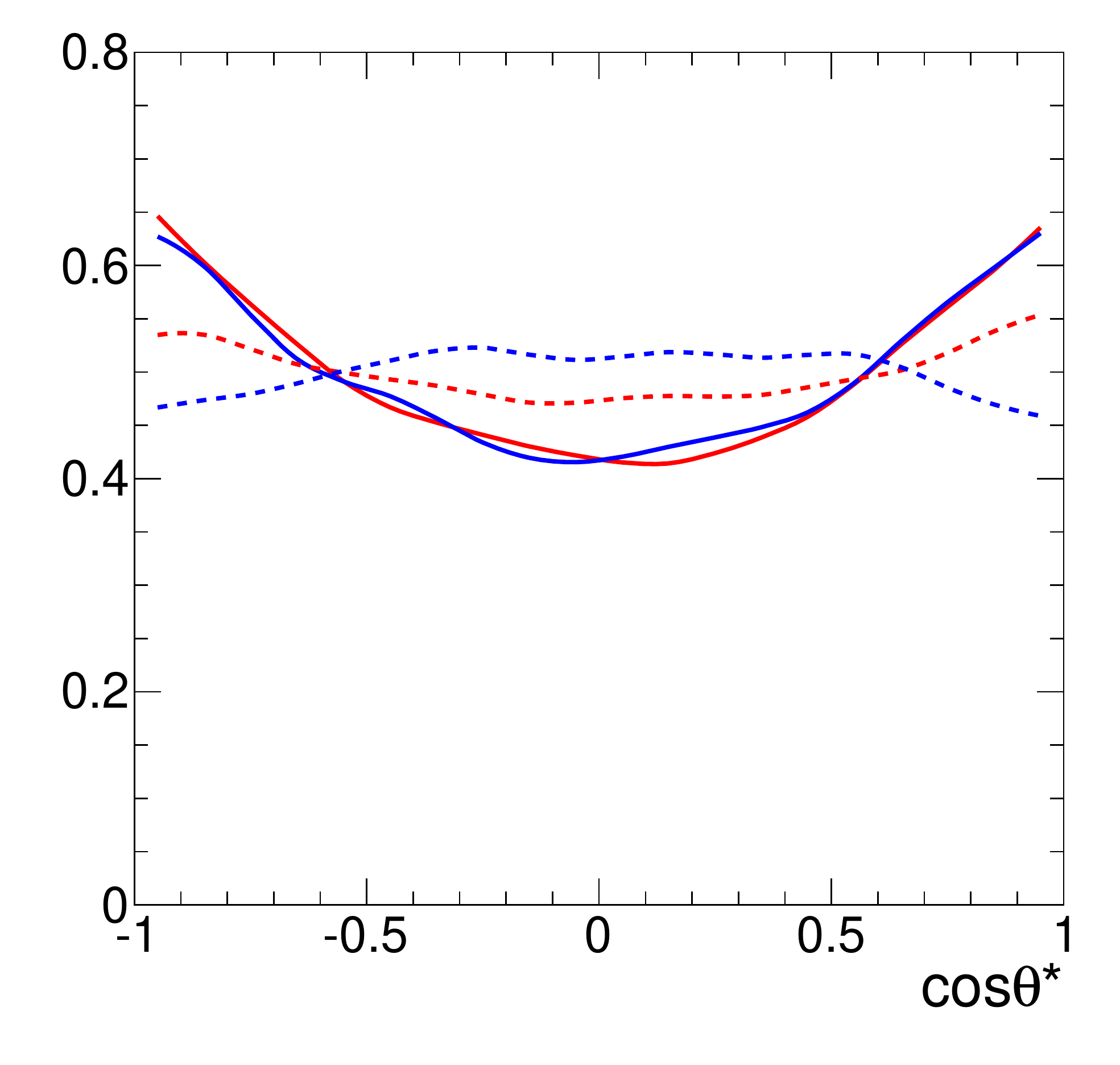}
 \includegraphics[width=0.275\textwidth]{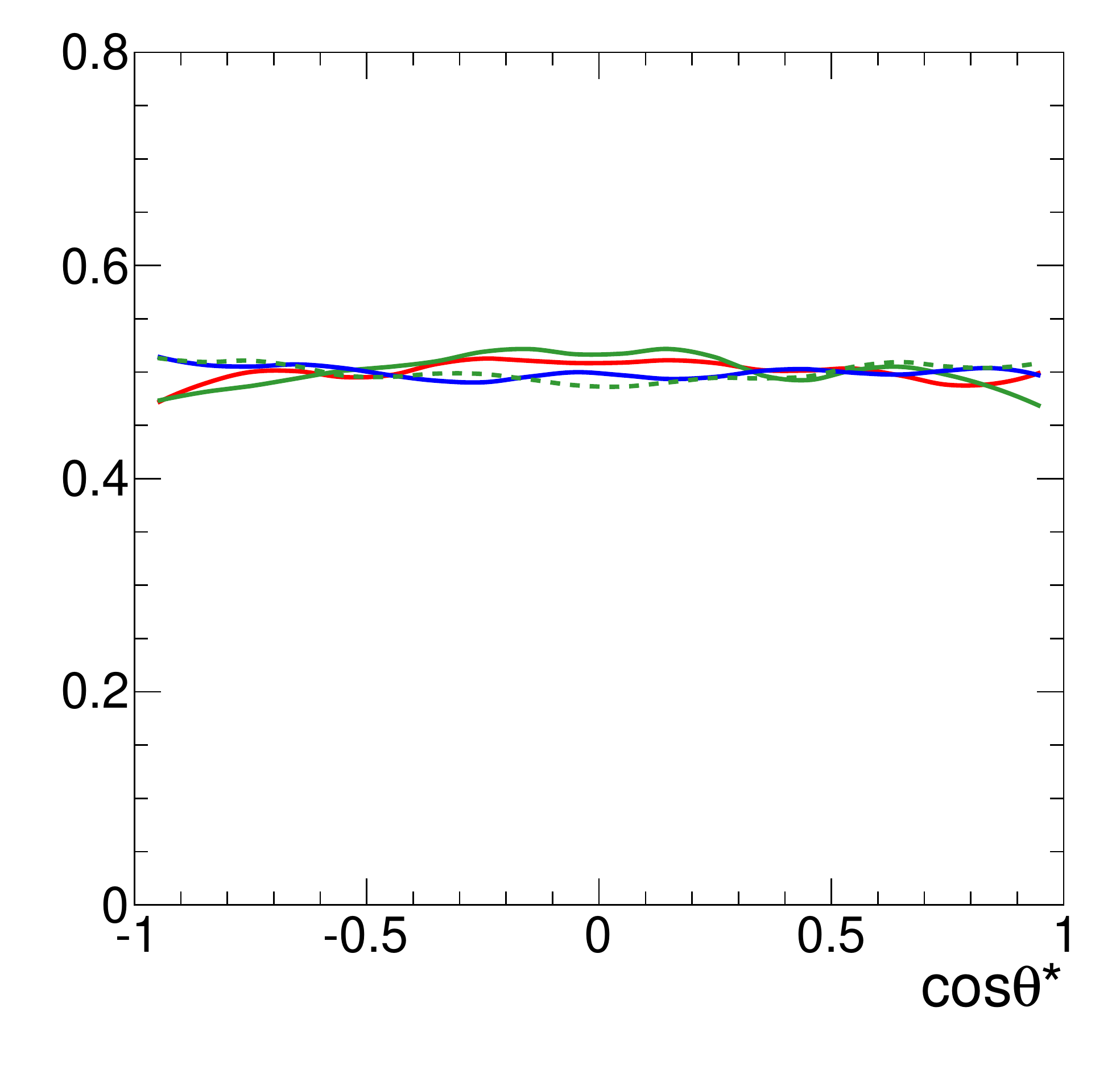} \\
 \includegraphics[width=0.275\textwidth]{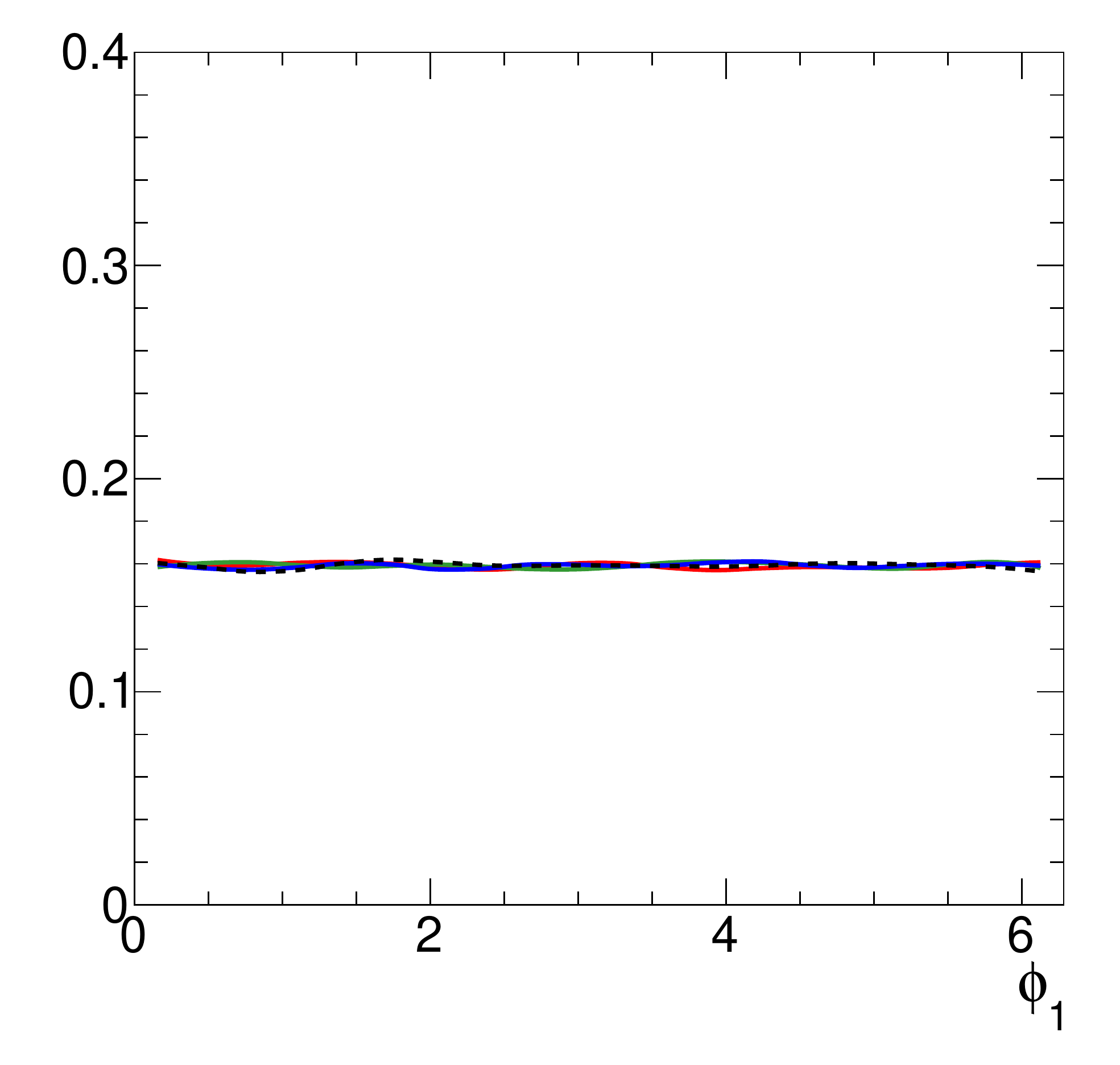}
 \includegraphics[width=0.275\textwidth]{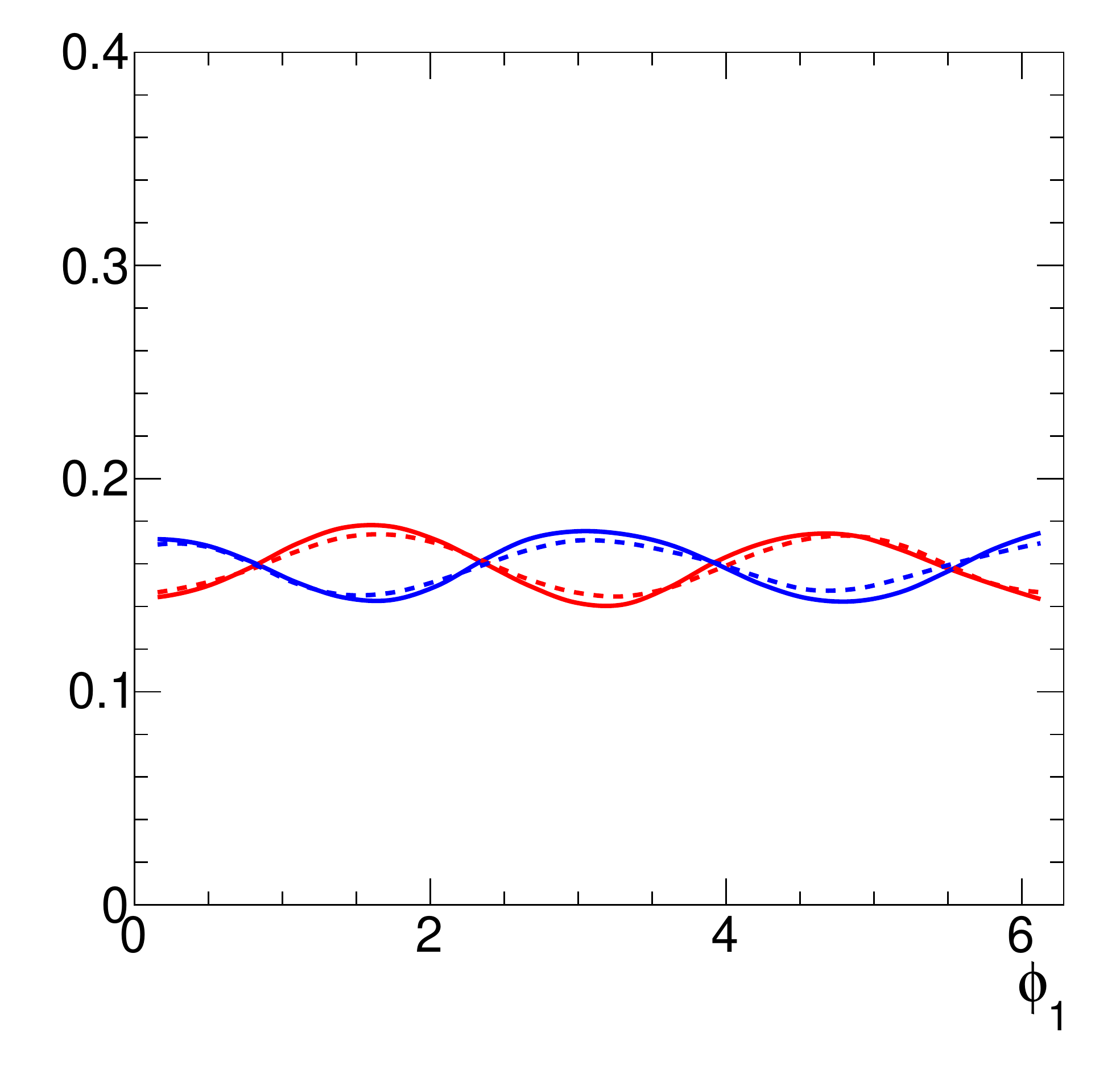}
 \includegraphics[width=0.275\textwidth]{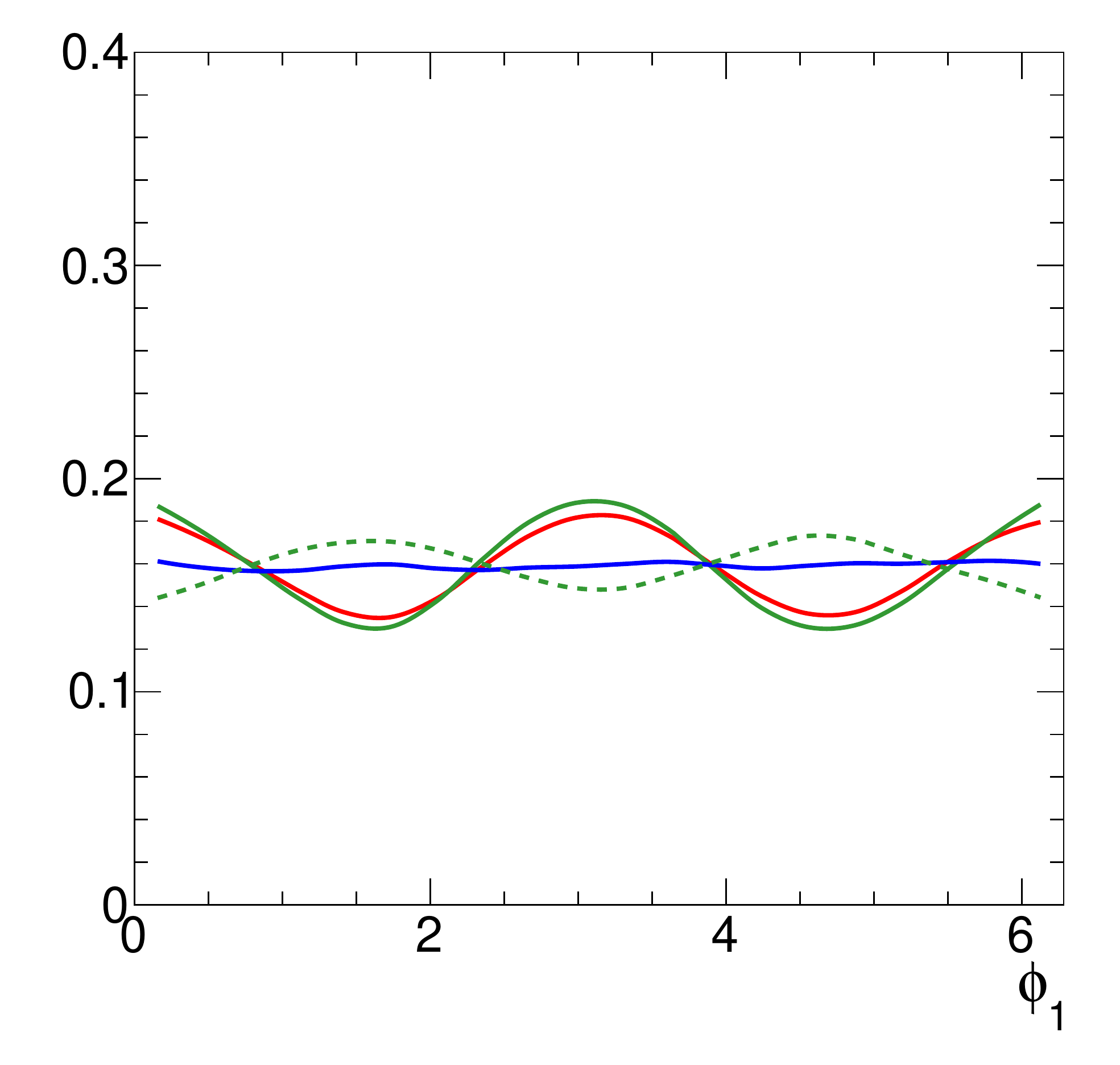} \\
 \includegraphics[width=0.275\textwidth]{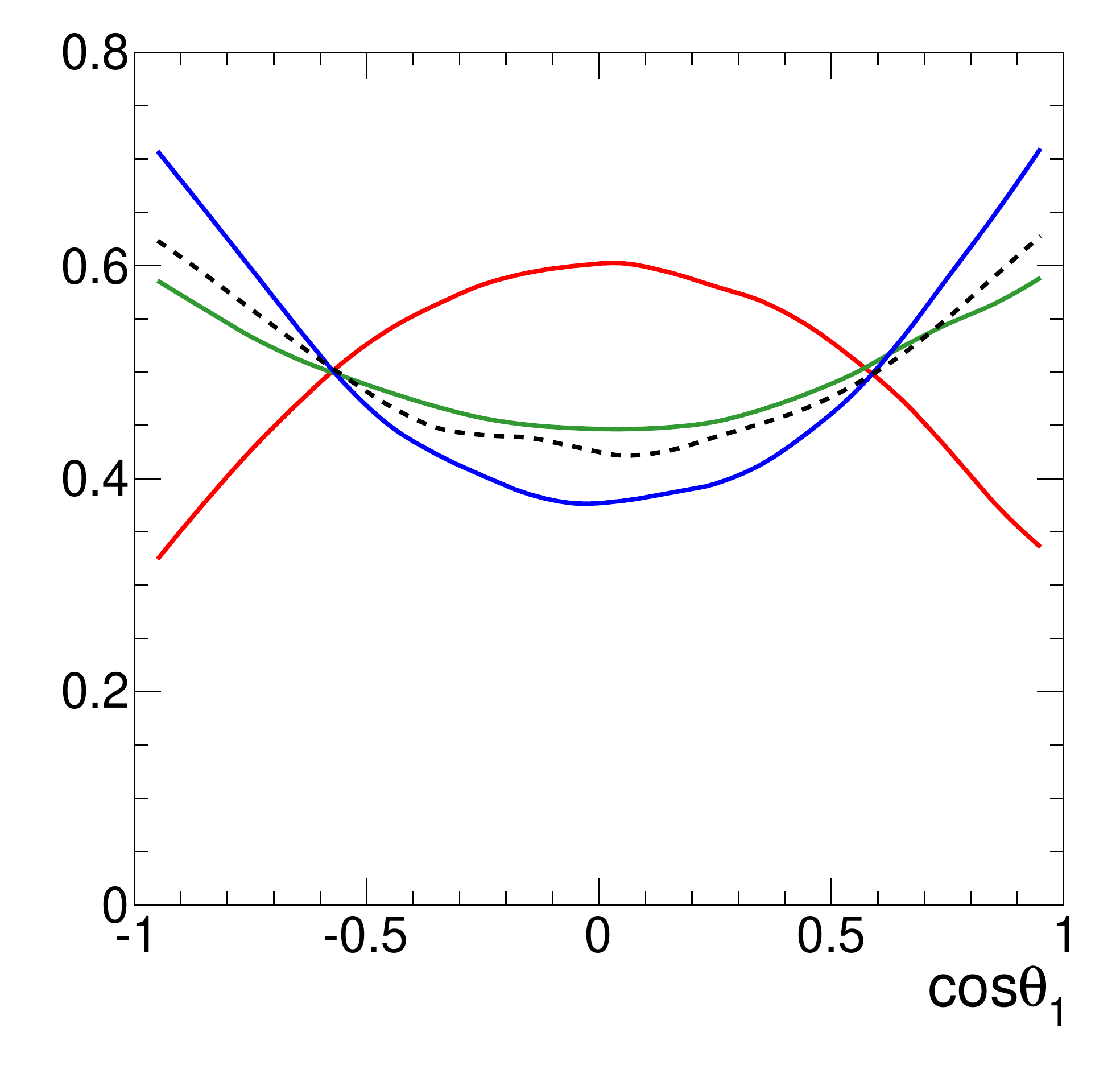}
 \includegraphics[width=0.275\textwidth]{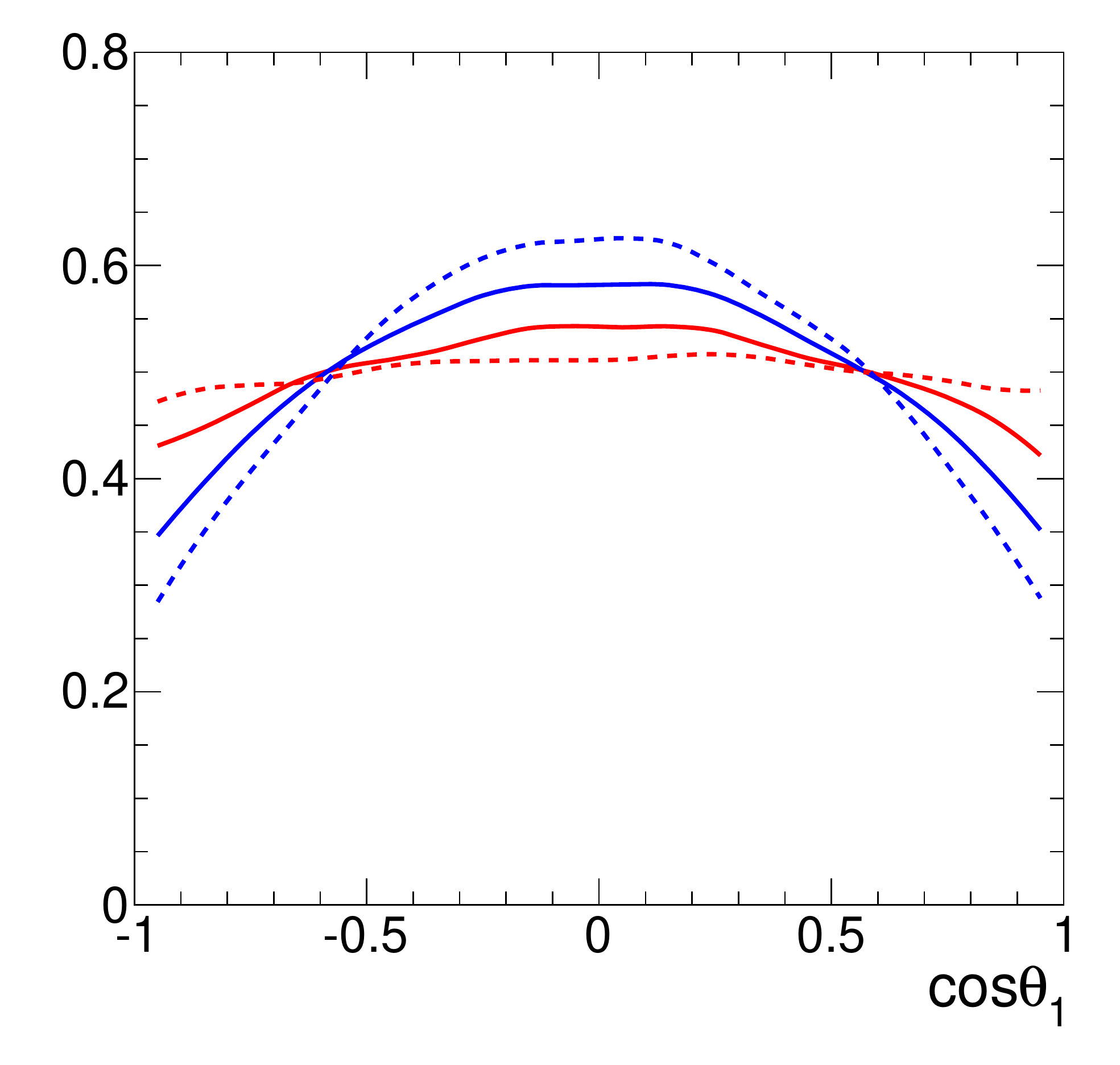}
 \includegraphics[width=0.275\textwidth]{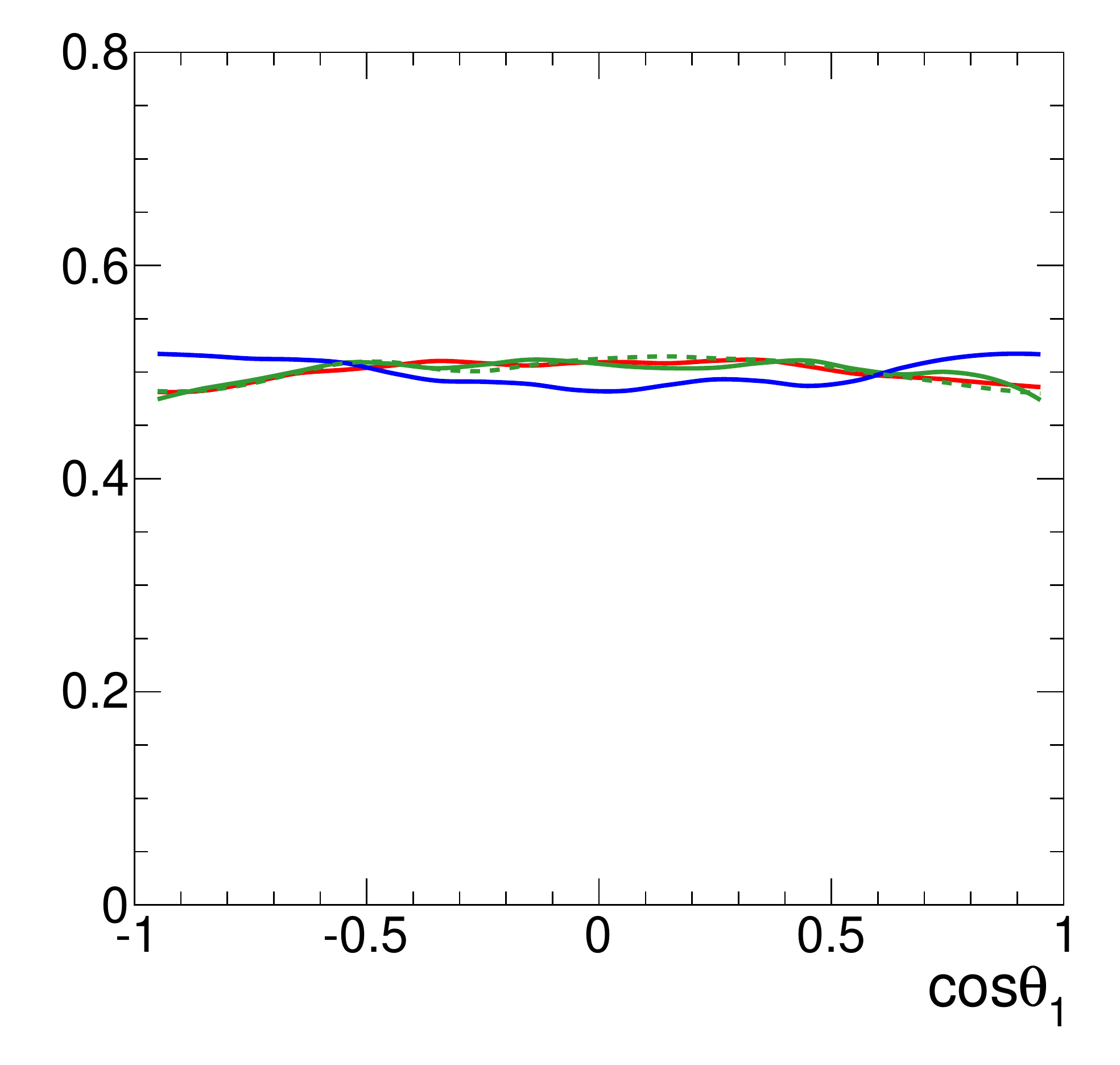} \\
 \includegraphics[width=0.275\textwidth]{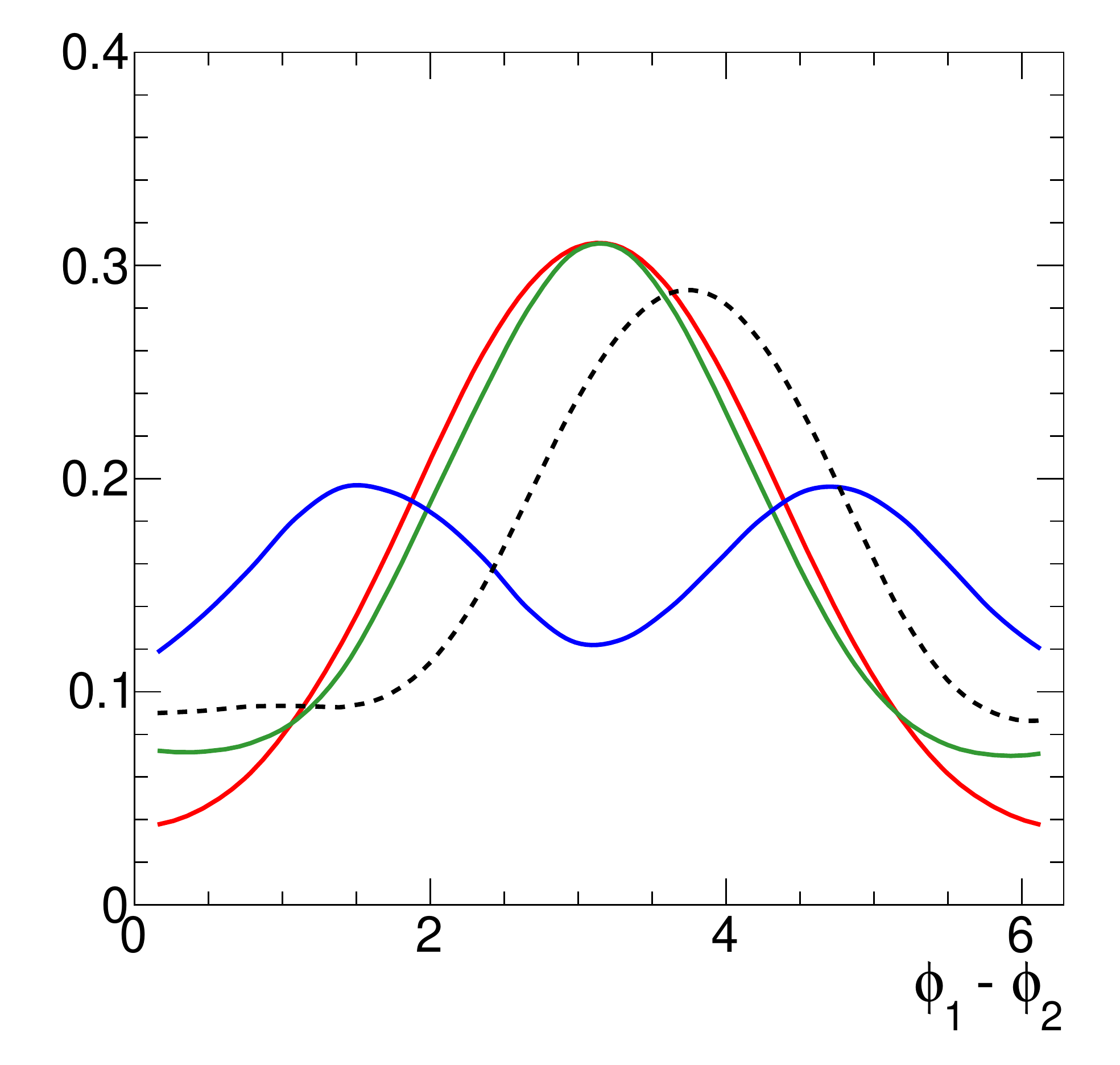}
 \includegraphics[width=0.275\textwidth]{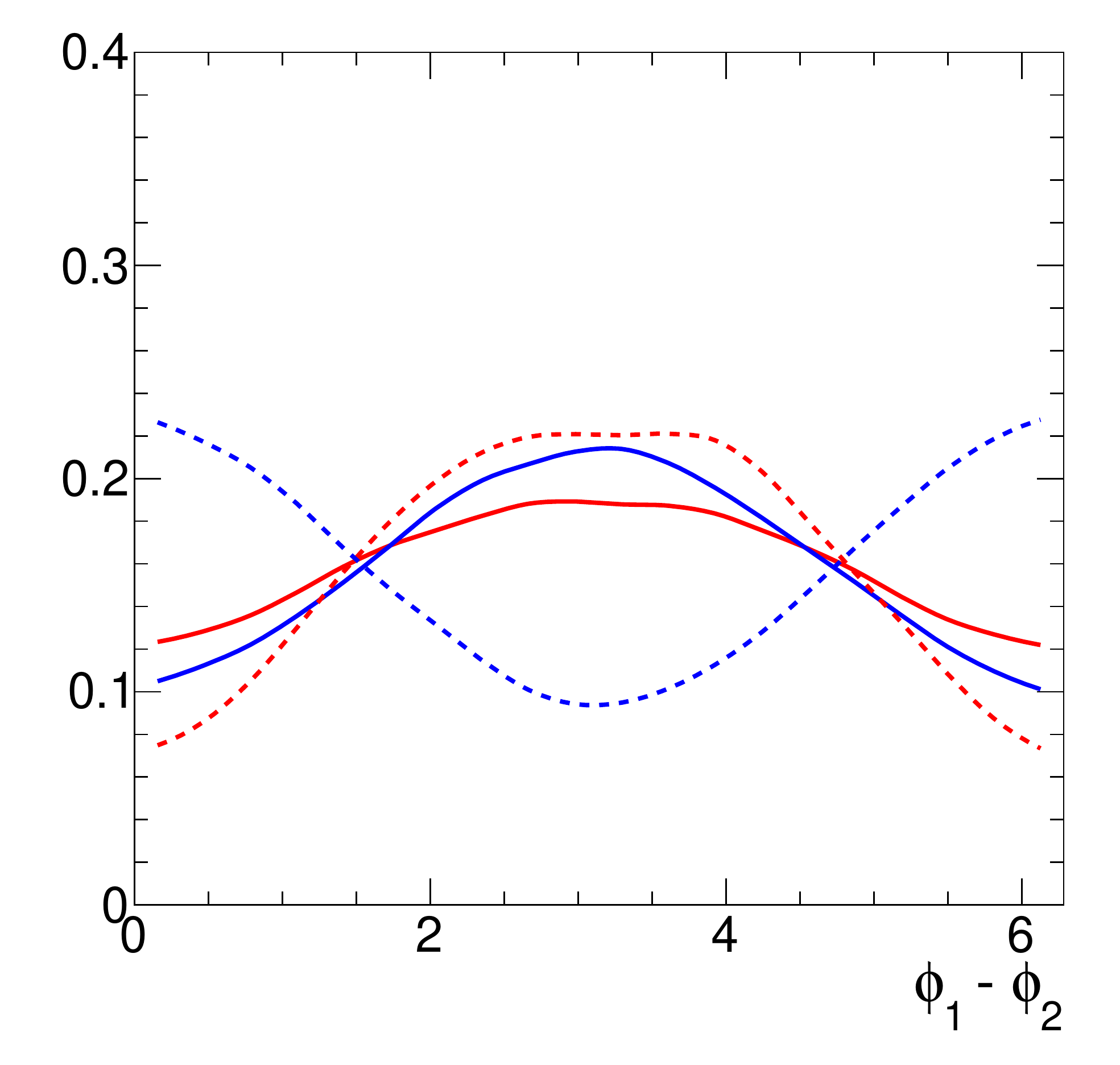}
 \includegraphics[width=0.275\textwidth]{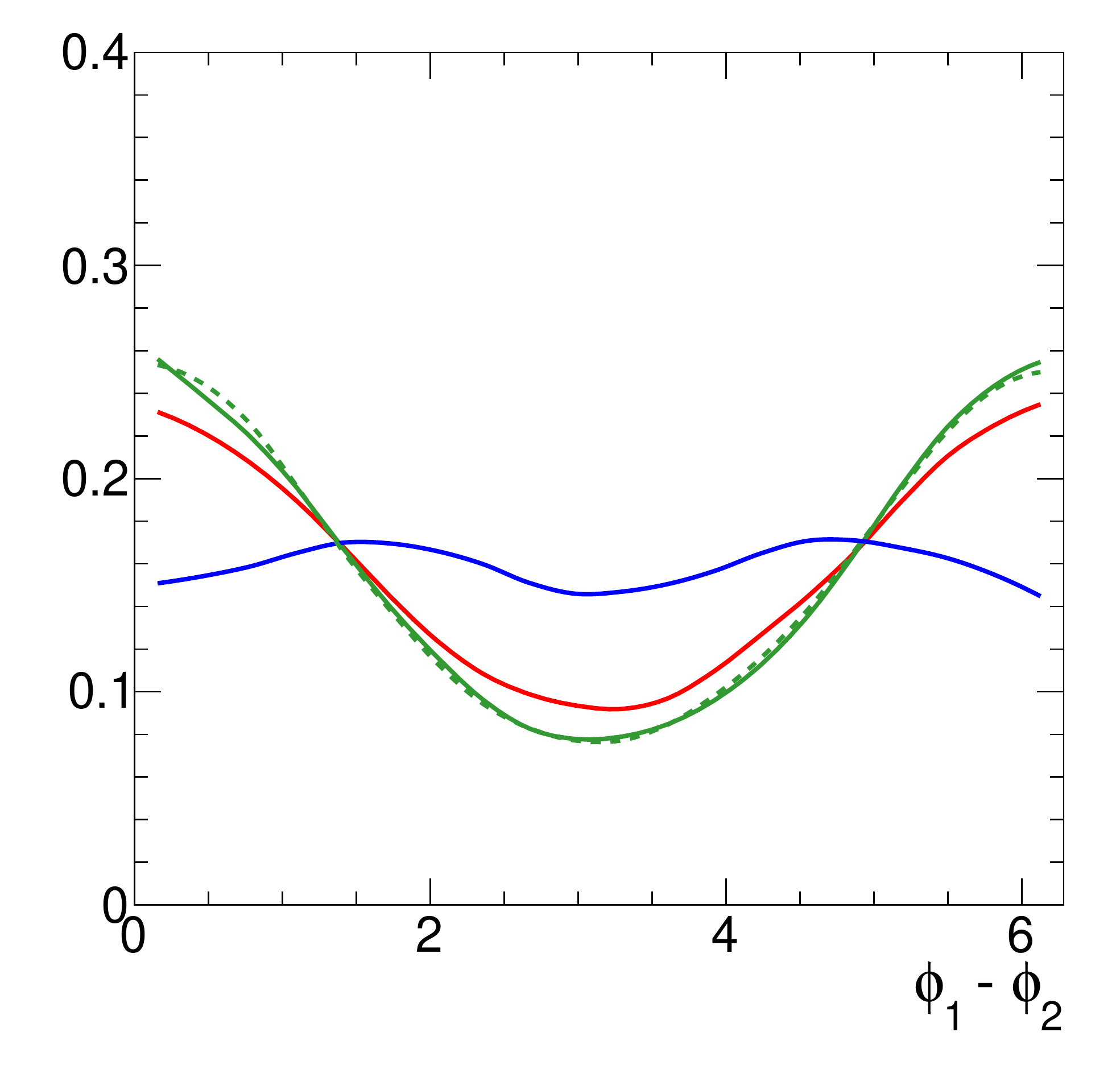} 
\caption{Mass and angular distributions in the $X\to WW$ analysis (cf. fig.~13 in
 the JHU paper~\cite{Bolognesi:2012mm}).}
\label{fig:WW}
\end{figure}

To validate our {\sc FeynRules} implementation, 
we show distributions in 
$pp\to X\to 4\ell$, comparing with the JHU results
in~\cite{Bolognesi:2012mm}.   
In table~\ref{tab:JHU} we give the choices of parameters to be made in 
order to obtain their benchmarks.
For all scenarios listed in that table one can see complete agreement in
the mass and angular distributions of the $X(J^P)$ decay products
in figs.~\ref{fig:ZZ_m}, \ref{fig:ZZ}, and~\ref{fig:WW}, where the
LO parton-level events were generated by {\sc MadGraph5}.

We note that our $CP$-even spin-0 parametrisation also includes the so-called
``derivative operators", that are absent in
the parametrisation of ref.~\cite{Bolognesi:2012mm}, and that give non-trivial
contributions to $X_0 \to VV$ decays~\cite{Artoisenet:2013puc}.
For spin~1 the $X_1VV$ interactions defined in
ref.~\cite{Bolognesi:2012mm} have one-to-one correspondence with the
$\kappa_{V_3}$ and $\kappa_{V_5}$ terms for both the $X_1WW$ and
$X_1ZZ$ cases.
We also note that a spin-2 state with non-universal
couplings to SM particles might have a very different behaviour with
respect to that of an RS-graviton, especially at high energies. 
See more details and non-trivial phenomenological implications due to
the NLO QCD effects in ref.~\cite{Artoisenet:2013puc}.

\begin{table}[h]
\center
\begin{small}
\caption{Parameter correspondence to the benchmark scenarios defined in Table
  I of ref.~\cite{Bolognesi:2012mm}. In each scenario, the $\kappa_i$ 
couplings  that are not explicitly mentioned are understood to be 
equal to zero.}
\begin{tabular}{cccc}
\hline
JHU scenario &\hspace*{5mm}& \multicolumn{2}{c}{HC parameter choice} \\
&& $X$ production & $X$ decay \\
\hline
$0^+_m$ && $\kappa_{Hgg}\ne0$ & $\kappa_{\rm SM}\ne0\ (c_{\alpha}=1)$\\
$0^+_h$ && $\kappa_{Hgg}\ne0$ & $\kappa_{H\gamma\gamma,HZZ,HWW}\ne0\ (c_{\alpha}=1)$\\
$0^-$ && $\kappa_{Agg}\ne0$ &  $\kappa_{A\gamma\gamma,AZZ,AWW}\ne0\ (c_{\alpha}=0)$\\
$1^+$ && $\kappa_{f_a,f_b}\ne0$ & $\kappa_{Z_5,W_5}\ne0\ $ \\
$1^-$ && $\kappa_{f_a,f_b}\ne0$ & $\kappa_{Z_3,W_3}\ne0\ $\\
$2^+_m$ && $\kappa_{g}\ne0\ $ & $\kappa_{\gamma,Z,W}\ne0\ $\\
\hline 
\end{tabular}
\label{tab:JHU}
\end{small}
\end{table}


%



\begin{small}


\bigskip 

\end{small}

\end{document}